\documentclass[fleqn,usenatbib]{mnras}

\usepackage{newtxtext,newtxmath}

\usepackage[T1]{fontenc}

\DeclareRobustCommand{\VAN}[3]{#2}
\let\VANthebibliography\thebibliography
\def\thebibliography{\DeclareRobustCommand{\VAN}[3]{##3}\VANthebibliography}

\usepackage{graphicx}	
\usepackage{amsmath}	
\usepackage{tabularx}
\usepackage{array}
\usepackage{multirow}
\usepackage{booktabs}
\usepackage{caption}
\usepackage{geometry}
\usepackage{morefloats}
\usepackage{fancyhdr}
\usepackage{supertabular}
\usepackage{multicol}
\usepackage{xcolor, soul}
\usepackage{multirow}
\usepackage{longtable}
\usepackage{rotating}
\usepackage{lscape}
\usepackage[justification=centering]{caption}
\usepackage{graphicx,subcaption} 
\usepackage{caption}
\usepackage{layout}
\usepackage{cancel}
\usepackage[many]{tcolorbox}
\newtcolorbox{cross}{blank,breakable,parbox=false,
	overlay={\draw[red,line width=5pt] (interior.south west)--(interior.north east);
		\draw[red,line width=5pt] (interior.north west)--(interior.south east);}}

\title[Long term Spectral Analysis of 3C\,279]{Understanding the Broadband Emission Process of 3C\,279 through Long term  Spectral Analysis}

\author[]{
	Aminabi Thekkoth,$^{1}$\thanks{E-mail: thekkothaminabi@gmail.com}
	S. Sahayanathan,$^{2,3}$\thanks{E-mail: sunder@barc.gov.in}
	Zahir Shah$^{4}$,
	Vaidehi S. Paliya$^{5}$
	and C. D. Ravikumar$^{1}$
	\\
	\\
	$^{1}$Department of Physics, University of Calicut, Malappuram, Kerala, India\\
	$^{2}$Astrophysical Sciences Division, Bhabha Atomic Research Centre, Mumbai-400085, India\\
	$^{3}$Homi Bhabha National Institute, Mumbai-400094, India\\
	$^{4}$Department of Physics, Central University of Kashmir, India\\
	$^{5}$Inter University Centre for Astronomy and Astrophysics, Ganeshkind, Pune, India
}

\begin{document}
	\label{firstpage}
	\pagerange{\pageref{firstpage}--\pageref{lastpage}}
	\maketitle
	
	\begin{abstract}
The long term broadband spectral study of Flat Spectrum Radio Quasars during different flux states has the potential 
to infer the emission mechanisms and the cause of spectral variations. To scrutinize this, we performed a detailed
broadband spectral analysis of 3C\,279 using simultaneous \textit{Swift}-XRT/UVOT and \textit{Fermi}-LAT observations 
spanning from August 2008 to June 2022.  We also supplement this with the simultaneous \emph{Nu}STAR observations
of the source. 
The optical/UV, X-ray, and $\gamma$-ray spectra were individually fitted 
by a power-law to study the long term variation in the flux and the spectral indices. 
A combined spectral fit of simultaneous optical/UV and X-ray spectra was also performed to obtain the transition energy at which the spectral 
energy distribution is minimum. 
The correlation analysis suggests that the long term spectral variations of the source are mainly associated with the variations in the low energy index and
the break energy of the broken power-law electron distribution which is responsible for the broadband emission.
The flux distribution of the source represents a log-normal variability while the $\gamma$-ray 
flux distribution showed a clear double log-normal behaviour. The spectral index distributions were again normal except for
$\gamma$-ray which showed a double-Gaussian behaviour. This indicates that the log-normal variability of the source 
may be associated with the normal variations in the spectral index. The broadband spectral fit of the source using synchrotron
and inverse Compton processes indicates different emission processes are active at optical/UV, X-ray, and $\gamma$-ray 
energies. 

	\end{abstract}

	
	
	
	\section{Introduction}
	\paragraph*{}
	The broadband spectrum of blazars is predominantly non-thermal in nature and extends 
	from radio-to-gamma-rays with many of them detected even at GeV/TeV 
        energies \citep{2007A&A...470..475A, 2011ApJ...740...64L, 10.1063/1.3076700}. 
	The opacity to high energy radiation along with its rapid flux variability suggests that  
	the emission is arising from a relativistic jet of plasma moving at an angle close to 
	the line of sight of the observer on earth \citep{1995MNRAS.273..583D}. Blazars are further
	classified into Flat Spectrum Radio Quasars (FSRQs) or BL Lacs depending upon
	the presence or absence of line features in their optical spectrum. Besides these
	line features, many FSRQs also exhibit strong thermal emission at UV and IR wavelengths.
	These thermal components are attributed to the emission from the accretion disk
	and the dusty environment of the AGN \citep{2009MNRAS.399.2041G,2007ApJ...669..884S}.
	
	The spectral energy distribution (SED) of blazars is characterized by two broad 
	peaks with the low energy component peaking at optical-to-X-ray energies and 
	the high energy component peaking at gamma-ray energies \citep{2010ApJ...716...30A}. The low energy component 
	is well understood to be the synchrotron emission arising from a relativistic distribution 
	of electrons losing their energy in the jet magnetic field \citep{2008MNRAS.387.1669G}. The high energy component
	is generally interpreted as inverse Compton scattering of low energy photons \citep{2000ApJ...545..107B}. The target
	photons for the inverse Compton scattering can be the synchrotron photon themselves,
	commonly referred  to as synchrotron self Compton (SSC), and/or the photon field external to the 
	jet, referred to as external Compton (EC). The plausible source for these external photons can be the 
	thermal emission from disk/dust and the line emission from the broad line emitting region \citep{1994ApJ...421..153S}.
	
	Blazars are categorized based on the photon energy at which the low energy synchrotron SED peaks.
	Accordingly, they are classified as low, intermediate, and high energy peaked blazars. Interestingly, the luminous blazars have low synchrotron peak energy and this anti correlation is termed  
	the blazar sequence \citep{2008MNRAS.387.1669G,1998MNRAS.299..433F}. FSRQs are the most luminous class of blazars and have the lowest synchrotron peak energies consistent with the blazar sequence. Another intriguing feature 
	of FSRQs is the `Compton Dominance' where the power radiated in high energy spectral 
	component is larger than that from the low energy synchrotron spectral component \citep{2013ApJ...763..134F,2017A&A...606A..44N}. 
	
	The broadband SED of high energy peaked blazars can be reasonably modelled under synchrotron 
	and SSC emission while for FSRQs, modelling the high energy component demands EC mechanism 
	in addition. Moreover, detailed spectral modelling of FSRQs suggests the X-ray emission be associated
	with the SSC emission process and the gamma-ray emission to the EC process \citep{2012MNRAS.419.1660S,2018A&A...616A..63A,2012MNRAS.424..789C, 2017MNRAS.470.3283S}. Two competing 
	scenarios for the EC process are the scattering of broad line photons (EC/BLR) or the thermal
	IR dust emission (EC/IR). However, the detection of GeV/TeV gamma-rays supports the EC/IR interpretation
	since the scattering of broad line emission to these energies is heavily suppressed by the 
	Klein-Nishina effects \citep{2013ApJ...771L...4C,2000ApJ...545..107B,2015ApJ...809..174D}. Identification of the target source photons for EC scattering
	also has the additional advantage of predicting the location of the emission region buried in the 
	blazar jet \citep{2021MNRAS.500.5297A,2013EPJWC..6105004J, 2015ApJ...803...15P}.
	
	3C\,279 (z=0.539; \citet{1965ApJ...142.1667L}) is one of the well-studied FSRQ and the first one 
	to be detected at GeV/TeV energies. The source was studied extensively in gamma-rays by 
	\textit{EGRET} during different flux states. With the advent of \textit{Fermi}, 3C\,279 has been
	monitored continuously at MeV-GeV energies and have witnessed series of extreme
	flaring events \citep{2016ApJ...824L..20A,2012ApJ...754..114H,2015ApJ...807...79H, 2015ApJ...808L..48P, Shah:2019lzg}. 
	The high energy SED of the source peaks at this energy and the spectrum
	is significantly curved \citep{2016ApJ...824L..20A}. 
	A log-parabolic function represents the spectral shape better with 
	considerable variation in the model parameters during flare \citep{2021MNRAS.504.1103R,2021arXiv210311149Z,2020MNRAS.492.3829L}. For instance, during 
	2018 flare, variation in the peak energy was positively correlated with the gamma-ray flux 
	showing a `bluer when brighter' trend \citep{Shah:2019lzg}. The spectral curvature 
	at the peak energy also showed a positive correlation with the flux.
	The soft X-ray and optical band, on the contrary,  follow a simple power law with relatively harder 
	X-ray spectral index \citep{2021arXiv210311149Z,2020ApJ...890..164P}. 
	During a flare the X-ray hardness increases with the flux showing a `harder when brighter' 
	behaviour \citep{2010ApJ...716..835A,2020ApJ...902....2Y}. 
	Besides these, studies using longterm multiwavelength lightcurves have shown that the gamma-ray and X-ray fluxes 
	were correlated in 3C\,279; nevertheless, no significant correlation has been observed between optical and gamma-ray 
	energy bands \citep{2020ApJ...890..164P, 2020MNRAS.492.3829L}.
	
One of the striking features of 3C\,279 is its flux variability measured even up to minute timescales \citep{2016ApJ...824L..20A,2015ApJ...808L..48P}.
Such rapid variability demands a very compact emission region located well within the broad line emitting
region of the quasar \citep{2013EPJWC..6105004J,2012MNRAS.420..604N}. However, the location of the emission region inferred from the broadband SED modelling 
contradicts this since the $\gamma$-ray emission is better interpreted as EC/IR process \citep{2009ApJ...704...38S, 2016ApJ...817...61P}. 
Besides, the frequent flaring episodes encountered from this source also raise the question of whether the source poses two definite flux states. 
Studying the long term flux distribution have the potential to provide clues to this.  In general, the flux distribution  of 3C\,279
supports a log normal behaviour suggesting the physical process responsible for the variability to be 
multiplicative \citep{2018Galax...6..135R,2022ApJ...927..214G, 2021MNRAS.504.5074S}. This can possibly indicate 
the coupling of the blazar jet with the accretion disk where the variable emission from the latter is well understood to 
be log normal \citep{2010LNP...794..203M}. Alternatively, a log normal flux
distribution can also be an outcome of the Gaussian fluctuation associated with the particle acceleration
timescale \citep{2018MNRAS.480L.116S,2020MNRAS.491.1934K}.

The SED of FSRQs, and 3C\,279 in particular, have been reasonably well understood through broadband 
spectral modelling using synchrotron, SSC, and EC emission 
processes \citep{2000ApJ...545..107B,2012MNRAS.419.1660S,2017MNRAS.470.3283S,Shah:2019lzg}. However, the temporal 
behaviour of the source at different energy bands in connection with these emission 
models has not been addressed in detail. Such a study, besides refining the emission 
models, has the potential to identify the origin of the flaring mechanism. 
In this work, we performed a detailed analysis of optical--X-ray--gamma-ray spectral behaviour of 3C\,279 spanning 
from 2008 to 2022. 
Since the spectra at these energies are interpreted by 
different emission processes, we carried over a correlation study between the best fit parameters of the 
individual spectral fittings to identify the plausible reasons for the observed flux variations.
Further, we also performed a statistical broadband spectral fit using synchrotron and inverse Compton emission processes for two epochs with simultaneous \emph{Swift}, \emph{Nu}STAR and \emph{Fermi} observations to validate the emission processes.
This paper is organized as follows: In the next section \S\ref{data}, we provide the details of the observations and 
data reduction, in section \S\ref{multiwave}, we present the multiwavelength analysis and the correlation studies, 
in section \S \ref{distribution}, we present our results on the multiwavelength flux and index distribution, and in section 
\S \ref{broadband}, we present the details of broadband spectral fit using synchrotron and inverse Compton emission processes. 
Throughout this work, we adapt a cosmology with $\Omega_m = 0.3$, $\Omega_\lambda = 0.7$ and $H_0 = 70$ km s$^{-1}$\,Mpc$^{-1}$.
\section{Observations and Data Reduction}
\label{data}
3C\,279 is one of the well-studied FSRQ with a wealth of observations at different energy bands. To study its spectral 
behaviour at optical/UV, X-ray, and $\gamma$-ray energy bands, we performed a detailed analysis of the source 
using longterm observations by \emph{Swift}-UVOT, \emph{Swift}-XRT, and \emph{Fermi}-LAT telescopes on board. 
We also supplement these with the simultaneous \emph{Nu}-STAR observations of the source at hard X-rays available for two epochs.
\subsection{\textit{Fermi}-LAT Observations}
\paragraph*{} 
The Fermi $\gamma$-ray telescope is an international space observatory that observes the cosmos in a wide range of $\gamma$-ray energy. Its primary instrument, 
Large Area Telescope (LAT), detects photons of energy from 20 MeV to 1 TeV through 
the pair production process. In this work, we have analyzed the \textit{Fermi}-LAT observation of the source spanning
from 2008 to 2022 (MJD 54693-59754) in the energy range 0.1-300 GeV. 
A $10^{\circ}$ circular region centered at RA=194.0415, \mbox{Dec= -5.7887} was chosen as Region of interest (ROI) to download 
the data\footnote{\url{https://Fermi.gsfc.nasa.gov/ssc/data/access/lat/12yr_catalog}}. 
For data reduction, we used the \textit{Fermitools} software version 2.2.0 and followed the procedures given by
\footnote{\url{https://Fermi.gsfc.nasa.gov/ssc/data/analysis/scitools/}}.
First, we performed the preferred selection cuts in the event data on region, time, energy interval, and type. For pass8 data, the recommended event 
class and type for studying point sources are \textit{evclass=128 and evtype=3}. The region and energy interval selections were kept the same as that of 
the downloaded observation. We also put a maximum cut on the zenith angle at $90^{\circ}$ to eliminate earth limb events. The `\textit{gtmktime}' tool has 
been used to update good time intervals based on the spacecraft parameters. The current GTI filter expression recommended is `(DATA\_QUAL==1)\&\&(LAT\_CONFIG==1)'. 
A livetime and exposure maps have been created using `\textit{gtltcube}' and `\textit{gtexpmap}' respectively. The live time is the accumulated time during which, 
LAT is actively recording the photons. `\textit{gtltcube}' integrates the livetime as a function of sky position and off-axis angle. A binned likelihood analysis 
method has been performed to fit the data over the whole time interval.
We used the standard templates recommended gll\_iem\_v07 and  iso\_P8R3\_SOURCE\_V3\_v1 for modelling galactic diffuse and extragalactic isotropic background 
emissions. All $\gamma$-ray sources within 20-degree radius circle from 
the center were considered while fitting and their spectral shapes have been adopted from the 4FGL catalog. 
The parameters of all sources lying within the ROI ($10^{\circ}$ circle) have been kept free while that of the sources outside ROI were fixed to their 
catalog values. The significance of the $\gamma$-ray detection from all the positions was estimated using the test statistic defined by, $TS=2\log (\mathcal{L})$. 
Since $\mathcal{L}$ is the ratio of maximum likelihood values with and without the source in a particular position, a larger TS indicates a higher 
probability for a source to be in the position. The spectral parameters of all sources for which TS<25 were then kept frozen. The output file thus obtained 
was used as the input sky model in further analysis.

We obtained longterm $\gamma$-ray lightcurves of 3C\,279 for two types of binning, a constant time interval (3 day) and adaptive binning. 
The adaptive binning is a new technique for the lightcurves, in 
which bin widths are estimated by setting a constant relative uncertainty on the flux \citep{2012A&A...544A...6L, 2022MNRAS.517.2757S}. Unlike the constant time bin method, 
which smooths out the minute scale variability, this method is more suitable for studying sources, especially in flaring periods. 
The adaptive time bins for a 20$\%$ constant relative flux density were obtained following the steps in the documentation 
\footnote{\url{https://www.slac.stanford.edu/~lott/ABM_mult_P8.tar.gz}}.
The optimum energy for estimating the adaptive time bins was computed as 157 MeV. 
The same was chosen as the minimum energy, $E_{min}$ for estimating fluxes in the adaptively binned lightcurve. 
The unbinned likelihood analysis has been adopted for obtaining lightcurves in the case of adaptive as well as 3 day binning criteria.
We chose the power-law2 function to model the $\gamma$-ray spectrum of 3C\,279 and used 
an iterative approach for the likelihood fit to obtain convergence in fit. Initially `DRMNFB' optimizer has been used and all the sources with TS<9 have been deleted once found no convergence in the trial. Additionally, all 
the parameters other than the norm for sources with TS between 9 and 50 had been kept frozen in such cases. All the final
fits have been optimized using `Newminuit' method.
\subsection{Swift Observations}
\paragraph*{}
In the X-ray and optical/UV bands, a total of 491 observations are available for 3C\,279 up to 59754. We obtained all the 
available Swift X-ray spectra using the automated online tool \textit{`Swift-XRT data products generator'}\footnote{\url{http://www.Swift.ac.uk/user objects/}}. 
This tool provides X-ray light curves, spectra, images, and positions of any point source in the Swift XRT field of view. The source and the background
regions are selected automatically based on the count rate. This tool also performs the corrections for instrumental artifacts such as bad pixels or pile 
up of photons in the CCD \citep{2009MNRAS.397.1177E}. 
All the obtained spectra were rebinned to 20 minimum counts in each energy bin and fitted with an absorbed power-law in XSPEC \citep{1996ASPC..101...17A}.
For absorption, the neutral hydrogen column density is chosen to be $n_{H}=2.24\times10^{20}\,cm^{-2}$ \citep{Pian_1999}. 
The unabsorbed integrated flux in the energy range of 0.3-10 keV and spectral index has been obtained from the best fit results. 
For the present work, we considered only those spectra which are well fitted with a power-law (reduced chi-square between 0.6 and 1.8). 
Additionally, We rejected some spectra due to very low exposure periods and a very low number of energy bins. After these reductions,
we were left with 326 X-ray spectra and the fit details are given in Table \ref{tab:xrt}.

All the available UVOT observations for 3C\,279 have been downloaded from the heasarc archive. This instrument has 6 filters out of which 3 
are at optical wavelengths (V, B and U) and the rest 3 are at UV wavelengths (UW1, UM2 and UW2). We followed the standard procedures of 
data reduction given by the tutorial\footnote{\url{https://Swift.gsfc.nasa.gov/analysis/threads/uvot_thread_spectra.html}} for obtaining spectral files. 
For each observation, the images over all the extensions were summed up using the \emph{uvotimsum} tool for every filter.
A circle of radius 6 arcsec centered at the source has been chosen to extract the source counts while, for the background estimation, 
we have used a circle of radius 20 arcsec in a source-free region near the target. For all the observations, the spectral products 
corresponding to each filter were then obtained using \textit{uvot2pha} tool. 
For the spectral fit, we considered only those \emph{UVOT} observations for which the images are available at least in 3 filters. The selected 
optical/UV spectra are then fitted using XSPEC with an absorbed power-law model. The integrated fluxes were corrected for galactic 
absorption by fixing the value of parameter E(B-V) magnitude to 0.025 \citep{2011ApJ...737..103S}. 
For some observations, we have included an additional 3\% systematic
error in order to obtain the reduced chi-square less than 2. Those observations with reduced chi-square larger than 2 even after adding 3\%
systematic error are not considered in the present work. Similarly, we also rejected those observations whose power-law spectral fit resulted 
in a reduced chi-square of less than 0.6. Finally, we are left with 189 \emph{UVOT} observations and the details are given in Table \ref{tab:uvot}.
 
For the combined optical/UV--X-ray spectral analysis, some of the \emph{Swift} observations are excluded since either UVOT or the 
XRT spectra were not available due to the selection criteria mentioned above. We were finally left with 170 \emph{Swift}
observations for which simultaneous optical--X-ray information was available. Similarly, for the simultaneous $\gamma$-ray 
analysis, we considered the adaptively binned \emph{Fermi} spectrum which overlaps (at least 
partially) with the \emph{Swift} observation epochs. This resulted in 260 simultaneous X-ray--$\gamma$-ray observations
and 164 optical/UV--$\gamma$-ray observations which are used in the present study. However, we have 
included all the selected observations for the spectral study of the individual energy band.
\begin{table}
	\centering
	\setlength \extrarowheight{3pt}
	\setlength{\tabcolsep}{2pt}
	\scriptsize
	\begin{tabular}{cccccc} 
		\hline
		\hline
		\textbf{Obs.Id}  & \textbf{Time in MJD} & \textbf{Flux} &  \textbf{spectral index, $\Gamma_{x}$} &   \textbf{$\chi^{2}$} & \textbf{Dof}\\
		& & \textbf{$10^{-11}$\,ergs\,cm$^{-2}$\,s$^{-1}$} & & &\\
		\hline
		\hline
		
		000 35019001 & 53748.0 & $ 1.498 \pm 0.057 $ & $ 1.517 \pm 0.04 $ & 111.22 & 132.0 \\
		000 35019002 & 53749.0 & $ 1.414 \pm 0.066 $ & $ 1.579 \pm 0.052 $ & 70.27 & 96.0 \\
		000 35019003 & 53751.2 & $ 1.636 \pm 0.202 $ & $ 1.397 \pm 0.13 $ & 16.06 & 15.0 \\
		000 35019004 & 53752.0 & $ 1.592 \pm 0.051 $ & $ 1.521 \pm 0.033 $ & 178.63 & 179.0 \\
		000 35019005 & 53753.1 & $ 1.766 \pm 0.061 $ & $ 1.545 \pm 0.037 $ & 160.48 & 148.0 \\
		\hline
		\hline
		\end{tabular}
		\caption{\small{Table showing results of power-law fitting of Swift-XRT observations. Col.\,1 \& 2 represents the Swift observation id and time in MJD of the observation. Col.\,2  represents the unabsorbed flux and 3 the spectral index respectively. Col.\,4 \& 5 gives value of chisquare and degrees of freedom of the fitting. The Entire tabular data is made available in machine readable form in the online journal.}}
		\label{tab:xrt}
		\end{table}
\subsection{NuSTAR Observations}
\paragraph*{}
We have used two \emph{Nu}STAR Observations (60002020002 and 60002020004) of 3C\,279 which are publicly available 
and  having simultaneous \emph{Swift} observations for the broadband SED analysis. The data were downloaded from the  
archive 
and reduced using standard pipeline techniques and software 
version v1.9.7 as described in \footnote{\url{https://heasarc.gsfc.nasa.gov/docs/nustar/analysis/}}. The \emph{nupipeline} tool was used 
to filter the event lists from the downloaded observations. Then the tool \emph{nuproducts} has been used to extract 
spectral products and response files for the two instruments FPMA and FPMB. To extract the source counts, we selected a 
circular region of 49 arcsecond radius centered at the source. Another circle of 60 arcsecond radius in a source-free region 
was selected as background region. Separate source and background region files were used for FPMA and FPMB 
observations. The obtained spectra were loaded in XSPEC and fitted with an absorbed power-law. 
The best fit photon power-law indices for the two observations are $1.73\pm0.02$ and $1.76\pm0.01$ with fit statistics $\chi^2$ (dof) as 529.42(528) and 678.77(737) respectively.
The unabsorbed fluxes in 3-79 keV energies is used for the broadband spectral study (section \S\ref{broadband}).
\begin{table}
	\setlength \extrarowheight{3pt}
	\setlength{\tabcolsep}{2pt}
	\scriptsize
	\begin{tabular}{cccccc} 
		\hline
		\hline
		\textbf{Obs.Id}  & \textbf{Time in MJD} & \textbf{Flux} &  \textbf{spectral index, $\Gamma_{\rm o/uv}$} &   \textbf{$\chi^{2}$} & \textbf{Dof}\\
		& & \textbf{$10^{-11}$\,ergs\,cm$^{-2}$\,s$^{-1}$} & & &\\
		\hline
		\hline
		000 30867007 & 54157.27 & $ 2.23 \pm 0.04 $ & $ 2.777 \pm 0.049 $ & 2.53 & 4.0 \\
		000 30867009 & 54266.79 & $ 1.07 \pm 0.02 $ & $ 2.665 \pm 0.046 $ & 5.12 & 4.0 \\
		000 30867017 & 54686.49 & $ 0.38 \pm 0.02 $ & $ 2.149 \pm 0.135 $ & 3.13 & 4.0 \\
		000 30867022 & 54695.6 & $ 0.45 \pm 0.02 $ & $ 2.14 \pm 0.114 $ & 4.65 & 4.0 \\
		000 30867024 & 54698.01 & $ 0.75 \pm 0.02 $ & $ 2.164 \pm 0.081 $ & 3.65 & 4.0 \\
		\hline
		\hline
		\end{tabular}
		\caption{\small{Table showing results of power-law fitting of Swift-UVOT selected observations. Col.\,1 \& 2 represents the Swift observation id and  MJD time of the observation. Col.2  gives the unabsorbed flux and Col.3 the spectral index. Col.\,4 \& 5 gives shows the chisquare and degrees of freedom respectively.This table is made completely available in machine readable form in the online journal.}}
		\label{tab:uvot}
		\end{table}

\begin{figure*}
	\centering
	\includegraphics[width=0.8\textwidth]{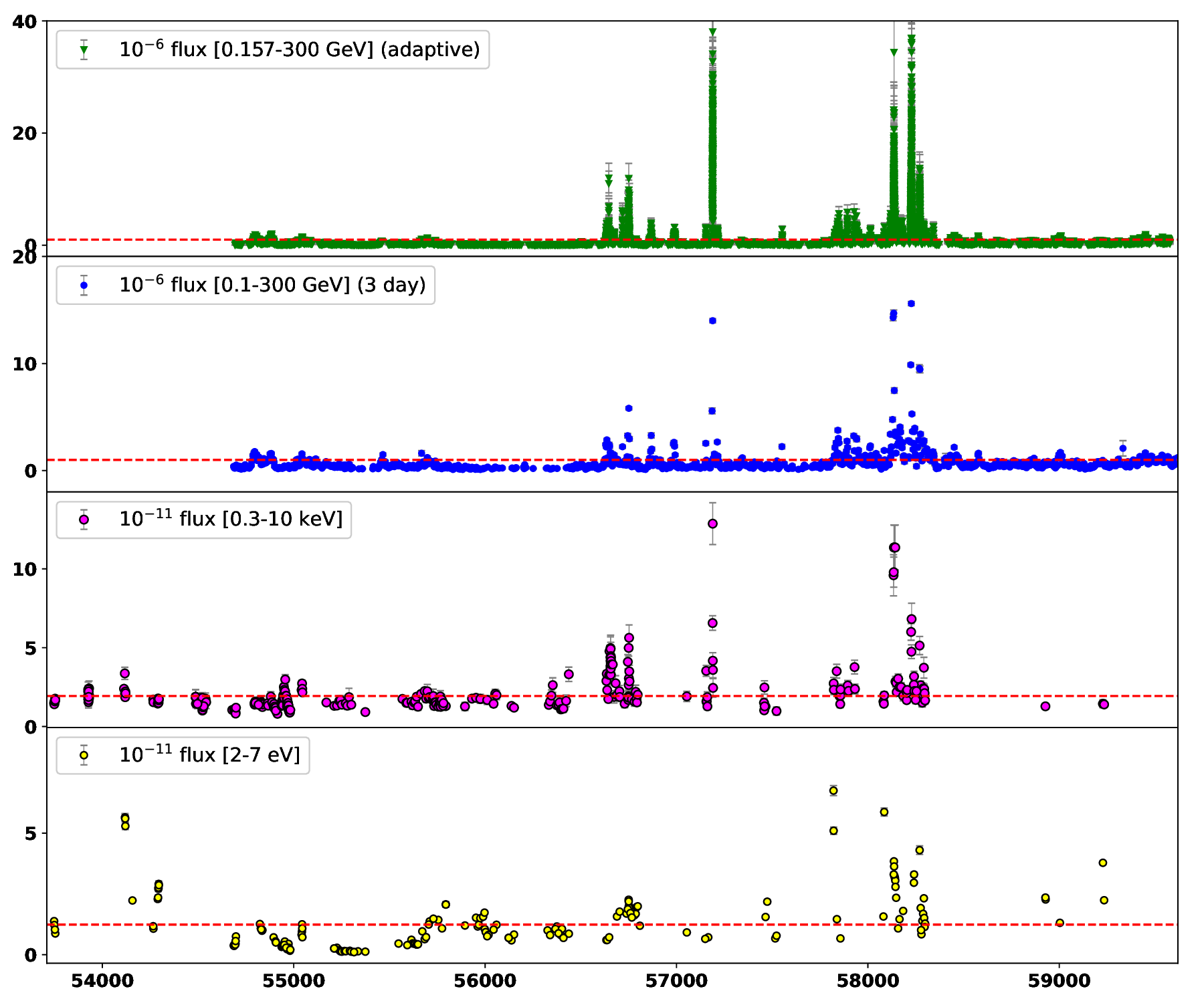}
	\caption{Plot of Multiwavelength light curves of flux in $\gamma$-ray, X-ray and optical energies of 3C\,279. The horizontal dashed lines represents the average flux. The X-ray and optical/UV fluxes are in units of \textbf{ergs}\,\textbf{cm}$^{-2}$\,\textbf{s}$^{-1}$ and $\gamma$-ray fluxes are in units of \textbf{phs}\,\textbf{cm}$^{-2}$\,\textbf{s}$^{-1}$. From top to bottom: panel 1 \& 2 displays the $\gamma$-ray light curves using Fermi observations in the energy range 0.1-300 GeV with adaptive and 3 day binning respectively. For adaptive binning, integrated fluxes were estimated in 0.157-300 GeV energy range. In the next panel, X-ray lightcurve is shown using Swift-XRT observations in the energy range 0.3-10 keV. In the bottom panel, optical/UV lightcurve is displayed using Swift-UVOT observations in the energy range 2-7 eV.}
	\label{lc}
\end{figure*}
\section{Multi-wavelength Analysis}
\label{multiwave}
\paragraph*{}
In Figure \ref{lc}, we show the multiwavelength light curve of the source in $\gamma$-ray, X-ray, and optical/UV bands.
It is evident from the multiwavelength lightcurve that the source exhibits significant variability in all energy bands and the instances 
where the flux enhancements are not correlated.
For example, we observe an optical/UV flux excess around the epochs 57827 and 59238; whereas, no significant flux enhancement is observed 
in X-ray and $\gamma$-ray bands. On the contrary, during epochs 56750 and 57188, flux enhancement is witnessed in $\gamma$-ray 
and X-ray energy bands with no such variation in optical band.
The simultaneous SED during correlated and uncorrelated flares for certain epochs were already studied under synchrotron and 
inverse Compton emission models \citep{2020MNRAS.498.5128R}. 
Here, we considered all the observations of the source by \emph{Swift} and \emph{Fermi} telescopes during its entire 
period of operation till 59754 MJD. The $\gamma$-ray light curves were obtained for two different binning criteria. The top 
panel corresponds to the adaptive binning technique, whereas, the second from the top is obtained with a constant binning of 3 days.
The average flux over the entire period is shown as a horizontal line in all the light curves. The value of the average optical/UV flux 
estimated is $1.24\times 10^{-11}$ $erg\, cm^{-2}\, s^{-1}$ and that for X-ray emission is $1.93\times 10^{-11}$ $erg\, cm^{-2} \,s^{-1}$. 
The average $\gamma$-ray flux for 3C\,279 was found to be $9.27\times\,10^{-7}$ $phs\,cm^{-2}\,s^{-1}$.

To investigate the relation between the optical, X-ray, and $\gamma$-ray energies, we studied the spectra at these energies using 
a power-law function. \emph{Swift} being a pointing telescope, the individual observation period of 3C\,279 spans mostly from 0.2 to 7 ks; whereas,
due to the scanning mode of operation, the \emph{Fermi}-LAT observations are continuous. The $\gamma$-ray spectra simultaneous to 
optical/UV and X-ray spectra were taken from the corresponding adaptive bins. The $\gamma$-ray spectrum of 3C\,279 is generally
curved and often modelled using a log-parabolic function. To verify whether the $\gamma$-ray spectrum from the selected bins (adaptive bins) show curvature,
we estimated the test statistics for power-law and log-parabola functions, TS(power-law) and TS(log-parabola). 
A large positive value of the difference, TS(log-parabola)- TS(power-law) will indicate significant curvature in the spectrum \citep{2020ApJ...890..164P}. 
Our study suggests the $\gamma$-ray spectra are well represented by a power-law and the details of this study are given in 
Table \ref{tab:ts_curv}. The integrated $\gamma$-ray flux is estimated by considering the best-fit power-law.
\begin{table*}
	\centering
\setlength\extrarowheight{3pt}
\setlength\tabcolsep{3pt}
\scriptsize
\begin{tabular}{llccccc} 
	\hline
	\hline
	\textbf{Tstart} & \textbf{Tstop} & \textbf{Flux} & \textbf{power-law Index}&  \textbf{Ts(power-law)}& \textbf{Ts(logparabola)} & \textbf{Ts(Curvature)} \\
	\textbf{MJD} & \textbf{MJD} & \textbf{10}$^{-7}$\textbf{phs} \textbf{cm}$^{-2}$ \textbf{s}$^{-1}$ & & & & \\
	\hline
	\hline
	54693.0 & 54696.0271 & $ 1.048 \pm 0.52 $ & $ 2.006 \pm 0.302 $ & 34.58 & 34.59 & 0.02 \\
	54693.0 & 54696.0271 & $ 1.048 \pm 0.52 $ & $ 2.006 \pm 0.302 $ & 34.58 & 34.59 & 0.02 \\
	54696.0271 & 54697.0139 & $ 6.201 \pm 1.743 $ & $ 2.391 \pm 0.273 $ & 57.42 & 57.42 & 0.0 \\
	54697.8665 & 54698.2684 & $ 8.895 \pm 2.604 $ & $ 2.13 \pm 0.235 $ & 72.46 & 72.46 & 0.0 \\
	54795.0751 & 54795.6055 & $ 15.33 \pm 3.844 $ & $ 2.199 \pm 0.208 $ & 110.21 & 110.21 & 0.0 \\
	\hline
	\hline
	\end{tabular}
	\caption{\small{Table showing Details of best fit parameter values of power-law fit of selected $\gamma$-ray observations (adaptive time bins) simultanoeus to Swift observtaions. Col.\,1  \& 2 gives the MJD tstart and tstop of the time bins. Col.\,3 represents the integrated flux estimated from power-law (0.157-300 GeV). Col.\,4 gives the power-law spectral index. The comparison of test statistics of power-law and logparabola model are given in Columns 5 to 7. Full table is available in machine readable form in the online journal.}}
	\label{tab:ts_curv}
	\end{table*}
\subsection{Spectral Transition} 
\label{trans}
\paragraph*{}
The broadband spectral modelling of 3C\,279 indicates that the optical/UV emission is by synchrotron process while the  X-ray emission 
is due to Inverse Compton scattering \citep{2012MNRAS.419.1660S,2015ApJ...808L..48P}. From the narrow band spectral fitting using power-law functions, we found that 
the optical/UV spectral index $\Gamma_{\rm o/uv} > 2$ and the X-ray spectral index $\Gamma_x < 2$ (Tables \ref{tab:xrt} \& \ref{tab:uvot}). 
This suggests the optical/UV emission fall on the decaying part of the synchrotron spectrum while the X-ray emission lie on the rising part of the IC 
spectrum \citep{2012MNRAS.419.1660S, 2020MNRAS.492.3829L}.
A combined spectral study can therefore probe the relation between these spectral components. Particularly studying 
the variation in transition energy, 
where the dominant emission shifts from synchrotron to IC, can investigate the enhancement of these components during different flux states. 
To facilitate this, we performed a combined spectral fit of optical/UV and X-ray observations using a broken power-law described as
\begin{equation}
	\label{eq:bknpl}
	\dfrac{df}{d\epsilon} = 
	\begin{cases}
		f_0 \,\epsilon^{-\Gamma_1} \quad &\textrm{for} \; \epsilon\,\leq \epsilon_{b}\\
		f_0 \,\epsilon_b^{\Gamma_2-\Gamma_1} \epsilon^{-\Gamma_2} \quad  &\textrm{for} \; \epsilon\geq \epsilon_{b}
	\end{cases}
\end{equation}
where $\Gamma_1$ and $\Gamma_2$ are the high and low-energy photon spectral indices, $\epsilon_{b}$ is break energy
and $f_0$ is the normalization. The parameter $\epsilon_{b}$ can be treated as the transition energy and to obtain
better constraints on it, we have fixed $\Gamma_1$ and $\Gamma_2$ to the best fit power-law indices of the corresponding 
optical/UV and X-ray spectrum. The best fit values of $\epsilon_{b}$ obtained from the simultaneous observations of \emph{Swift} UVOT 
and XRT are given in Table \ref{tab:bkn}, and range from $\sim$ 0.02 to 1.0 keV. 

\begin{figure*}
	\begin{subfigure}{0.46\textwidth}
		\includegraphics[width=0.8\textwidth]{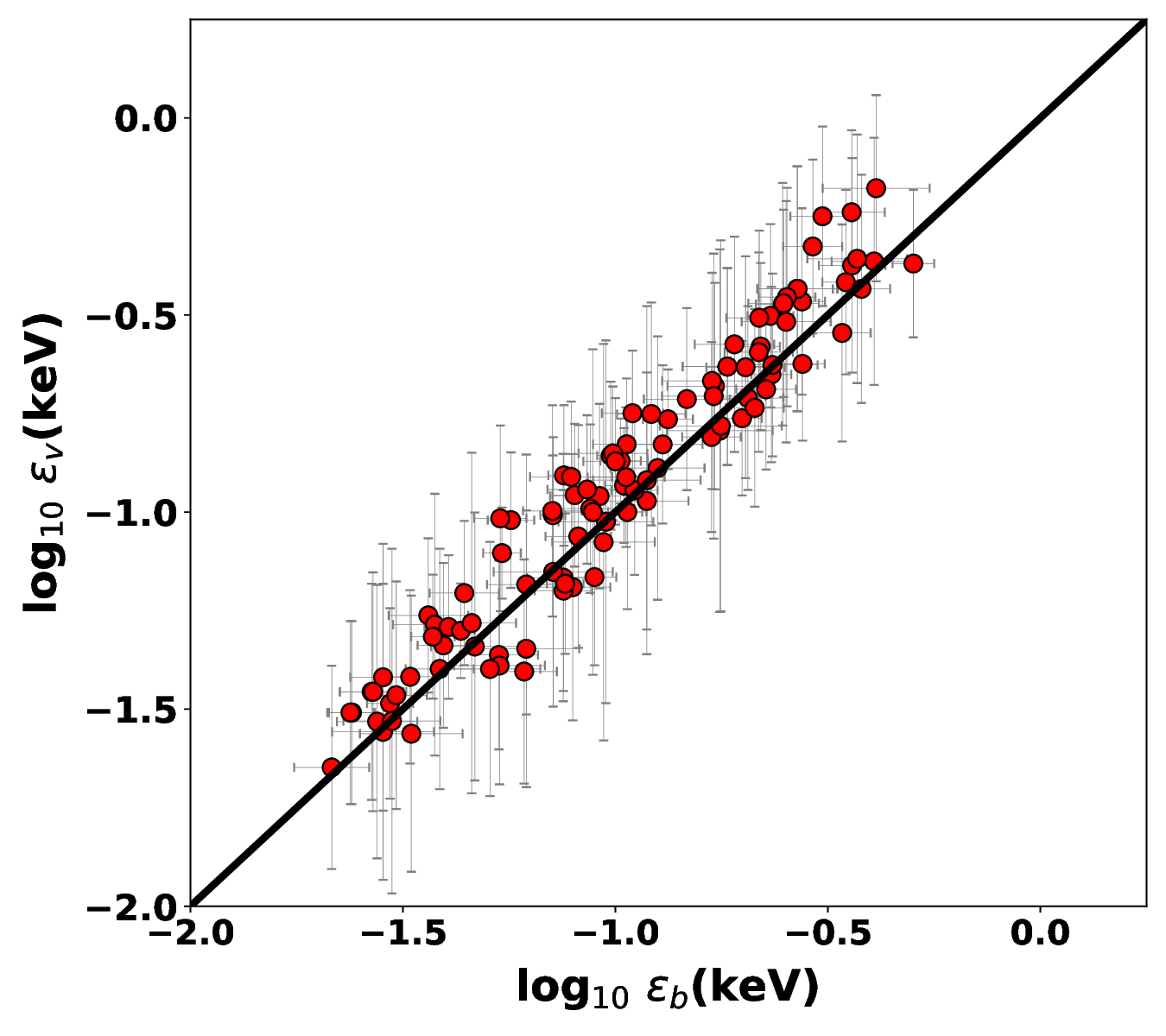}
		\caption{}
	\end{subfigure}
	\hfill
	\begin{subfigure}{0.46\textwidth}
		\includegraphics[width=0.8\textwidth]{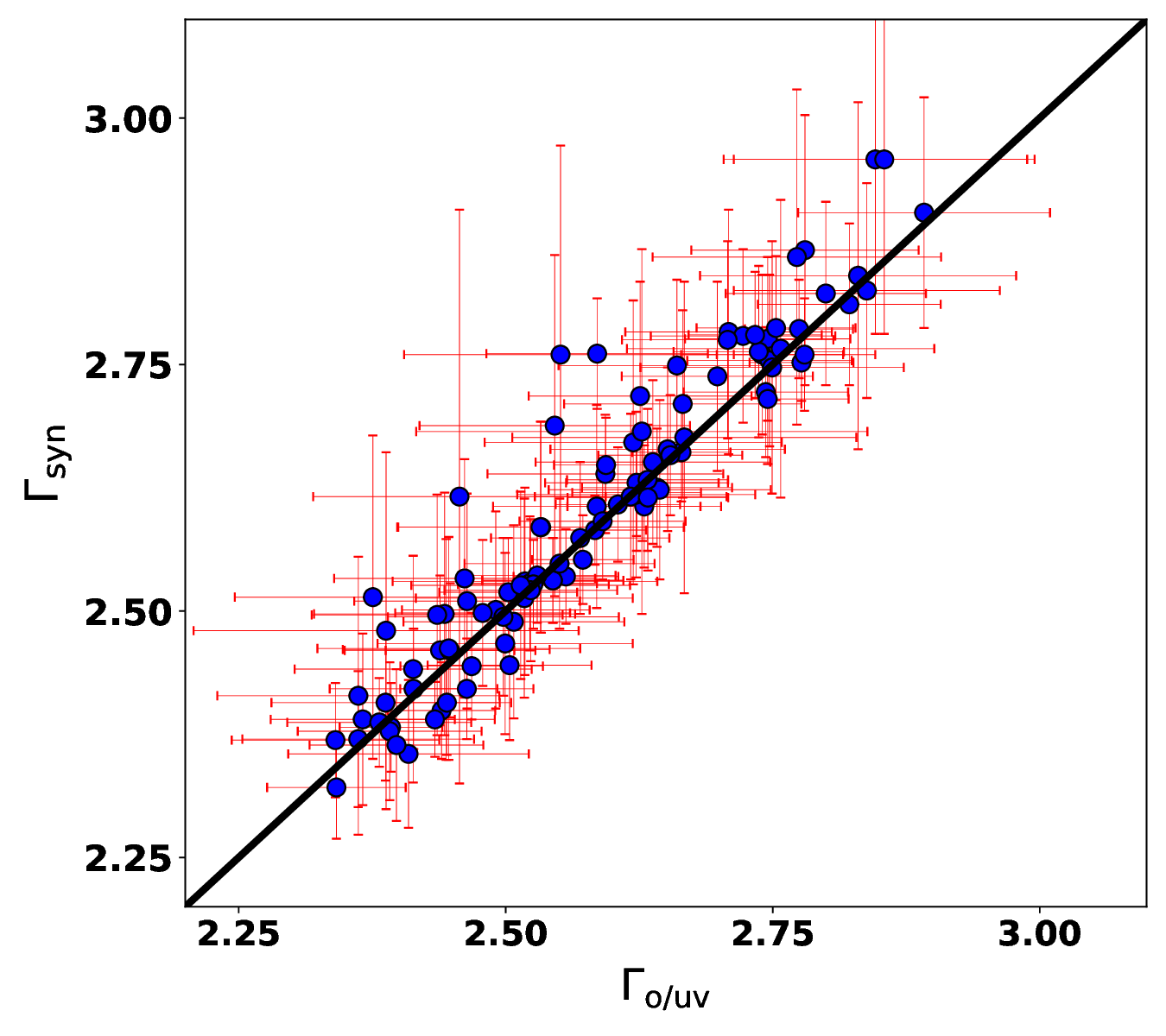}
		\caption{}
	\end{subfigure}
	
	\begin{subfigure}{0.46\textwidth}
		\includegraphics[width=0.8\textwidth]{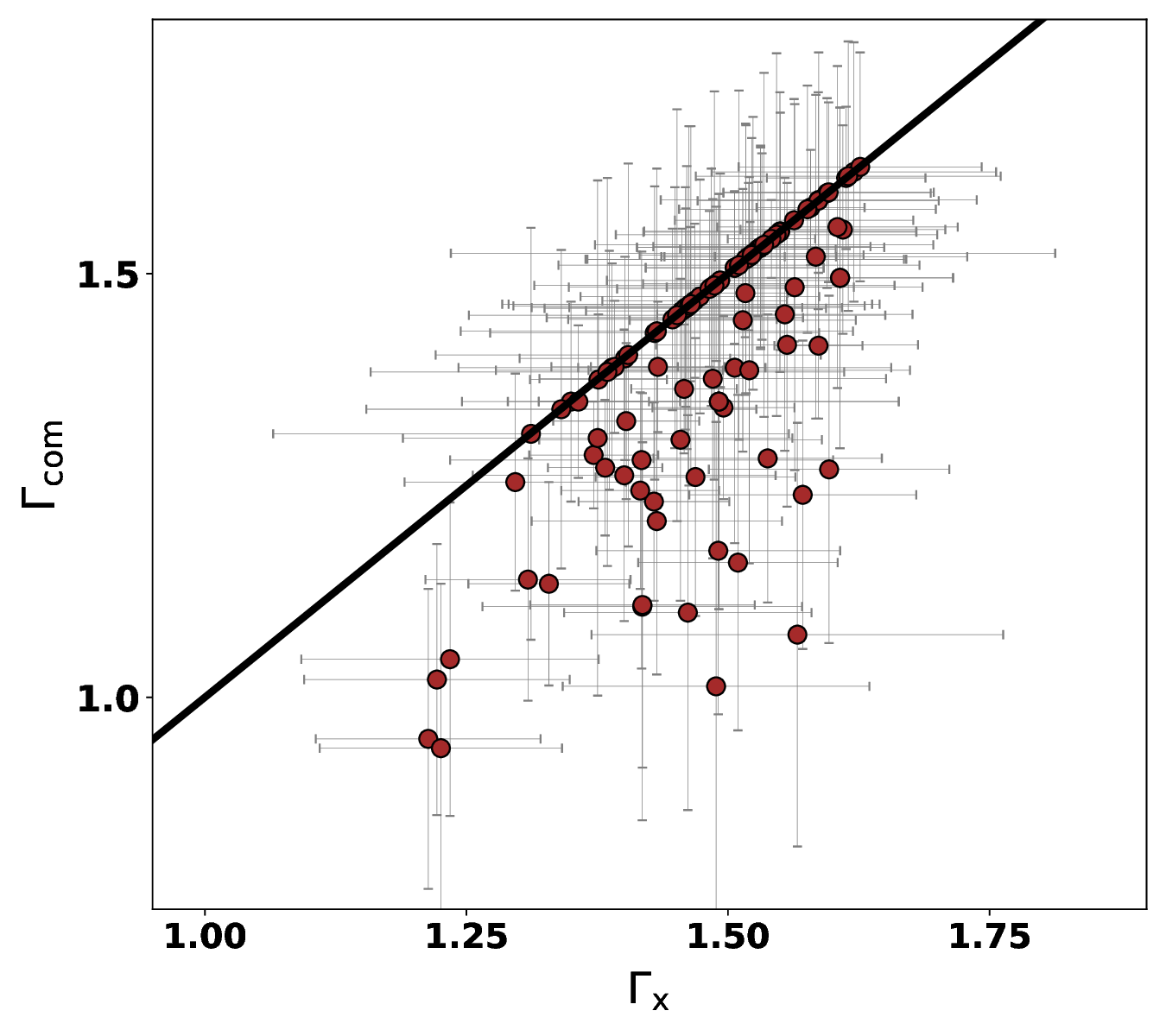}
		\caption{}
	\end{subfigure}
	
	\caption{Scatter Plots between the parameters obtained from broken power-law/power-law and double power-law fitting along with the identity line (black solid line). 
	(a): Plot between two estimates of transition energy, $\epsilon_b$ and $\epsilon_v$. (b): Plot between power-law optical/UV photon index, 
	$\Gamma_{\rm o/uv}$ and double power-law index, $\Gamma_{\rm syn}$. (c): Plot between two estimates of X-ray index from power-law and double power-law 
	fitting,$\Gamma_{x}$ and $\Gamma_{\rm com}$ respectively.}
\label{fig:eb-ev}
	\end{figure*}

A better representation of the transition energy can be obtained by fitting the simultaneous optical/UV 
and the X-ray data with a double power-law function. If we consider the underlying 
electron distribution responsible for the broadband SED of 3C\,279 to be a broken power-law, then
the optical/UV emission can be attributed to the synchrotron emission from high energy electrons and 
the X-ray emission to the inverse Compton scattering from the low energy electrons. We define the double power-law 
function as 
\begin{align}\label{eq:dpow}
	\dfrac{df}{d\epsilon}=f_{\rm 0}\left[ \left(\frac{\epsilon}{\epsilon_{\rm m}}\right)^{-\Gamma_{\rm syn}} 
	+ \left(\frac{\epsilon}{\epsilon_{\rm m}}\right)^{\Gamma_{\rm com}}\right]
\end{align}
Here, $\epsilon_{m}$ is the energy at which the synchrotron and IC fluxes are equal to $f_{0}$ 
while $-\Gamma_{\rm syn}$  and $\Gamma_{\rm com}$ are the
synchrotron and IC spectral indices, respectively. The 
valley energy ($\epsilon_v$) corresponding to minimum flux in SED will be 
\begin{equation}
	\epsilon_v = \epsilon_{m} \left(\frac{\Gamma_{\rm syn}}{\Gamma_{\rm com}}\right)^{{1}/\left({\Gamma_{\rm syn}+\Gamma_{\rm com}}\right)}
\end{equation}
and this can be treated as the transition energy.
The double power-law model is added as a local model in XSPEC \citep{2021MNRAS.506.3996J} with $f_{0}$, 
$\Gamma_{\rm syn}$, $\Gamma_{\rm com}$ and $\epsilon_v$ as the free parameters, and the combined spectral fits 
to optical/UV and X-ray data were performed. The results of the spectral fit are given in Table \ref{tab:dpl}.
In Figure \ref{fig:eb-ev} (a), we plot the transition energies obtained through a broken power-law ($\epsilon_b$) and double power-law 
($\epsilon_v$) functions along with the identity line. We find the estimate of the transition energy obtained using 
either methods closely match. We also compare the optical/UV and X-ray power-law spectral indices with $\Gamma_{\rm syn}$
and $\Gamma_{\rm com}$ in Figures \ref{fig:eb-ev} (b) and (c), along with the identity line. It is evident from the figures that 
$\Gamma_{\rm syn}$ and $\Gamma_{\rm o/uv}$ reasonably matches; whereas, $\Gamma_{\rm com}$ is relatively harder than 
$\Gamma_{x}$.
\begin{table}
\centering
\setlength\extrarowheight{3pt}
\scriptsize
\begin{tabular}{ccccc} 
\hline
\hline
\textbf{Obs Id} & \textbf{Time} &   \textbf{Break energy,$\epsilon_b$ } &  \textbf{$\chi^{2}$}& \textbf{Dof} \\
 & MJD    &keV &  & \\
\hline	
\hline
000 30867009 & 54266.79 &  $ 0.1029 \pm 0.0049 $ &  40.09 & 50.0 \\
000 30867022 & 54695.6 &  $ 0.757 \pm 0.1657 $ &  21.44 & 19.0 \\
000 30867024 & 54698.01 & $ 0.4083 \pm 0.1346 $ &  21.53 & 21.0 \\
000 30867028 & 54828.2  & $ 0.0987 \pm 0.008 $ &  13.62 & 23.0 \\
000 30867031 & 54834.23  &$ 0.0787 \pm 0.0076 $ &  13.82 & 18.0 \\
\hline
\hline
\end{tabular}
\caption{\small{Table showing Details of Combined spectral fitting of XRT and UVOT spectra with the Broken power-law model. Col.1 \& 2 gives the Swift observation id and time in MJD respectively. Col.3 shows the best fit value of break energy, $\epsilon_b$. Col.4 \& 5 is the values of chisquare and Dof. The complete table is available as online supplementary material.}}
\label{tab:bkn}
\end{table}

\begin{table*}
\centering
\setlength\extrarowheight{3pt}
\setlength\tabcolsep{3pt}
\scriptsize
\begin{tabular}{ccccccc} 
\hline
\hline
\textbf{Obs Id} & \textbf{Time} & \textbf{$\Gamma_{\rm syn}$\,($\Gamma_{\rm o/uv}$)} & \textbf{$\Gamma_{\rm com}$\,($\Gamma_{\rm x}$)}&  \textbf{Transition energy, E}$_v$& \textbf{$\chi^{2}$}& \textbf{Dof} \\
& MJD &  & &keV & & \\
\hline	
\hline
000 30867001 & 54113.0 & $ 2.761 \pm 0.056 $ & $ 1.415 \pm 0.086 $ & $ 0.4272 \pm 0.08 $ & 154.53 & 138.0 \\
000 30867003 & 54118.22 & $ 2.606 \pm 0.062 $ & $ 1.269 \pm 0.205 $ & $ 0.563 \pm 0.1282 $ & 34.61 & 27.0 \\
000 30867004 & 54119.22 & $ 2.63 \pm 0.059 $ & $ 1.239 \pm 0.182 $ & $ 0.5765 \pm 0.12 $ & 36.05 & 25.0 \\
000 30867005 & 54120.18 & $ 2.625 \pm 0.06 $ & $ 1.282 \pm 0.17 $ & $ 0.472 \pm 0.104 $ & 25.4 & 28.0 \\
000 30867007 & 54157.27 & $ 2.752 \pm 0.039 $ & $ 1.386 \pm 0.141 $ & $ 0.172 \pm 0.0217 $ & 11.07 & 23.0 \\
000 30867008 & 54265.78 & $ 2.763 \pm 0.087 $ & $ 1.613 \pm 0.084 $ & $ 0.0954 \pm 0.0164 $ & 54.0 & 45.0 \\
000 30867009 & 54266.79 & $ 2.661 \pm 0.05 $ & $ 1.488 \pm 0.091 $ & $ 0.1348 \pm 0.0145 $ & 44.72 & 48.0 \\
\hline
\hline
\end{tabular}
\caption{Table showing Details of Combined spectral fitting of XRT and UVOT spectra with the Double power-law model:Col.3 \& 4 represents the spectral index of optical and X-ray spectrum respectively. Col.\,5 gives the best fit values of Transition energy Energy, $\epsilon_{v}$. Col.6 \& 7 gives the values of chisquare and Dof. This table is made available in it's full form as online supplementary material.}
\label{tab:dpl}
\end{table*}
\subsection{Correlation Analysis}
\label{correlations}
\subsubsection{Flux-index Correlation}
\label{corre1}
\begin{figure*}
	\begin{subfigure}{0.48\textwidth}
		\includegraphics[width=0.85\textwidth]{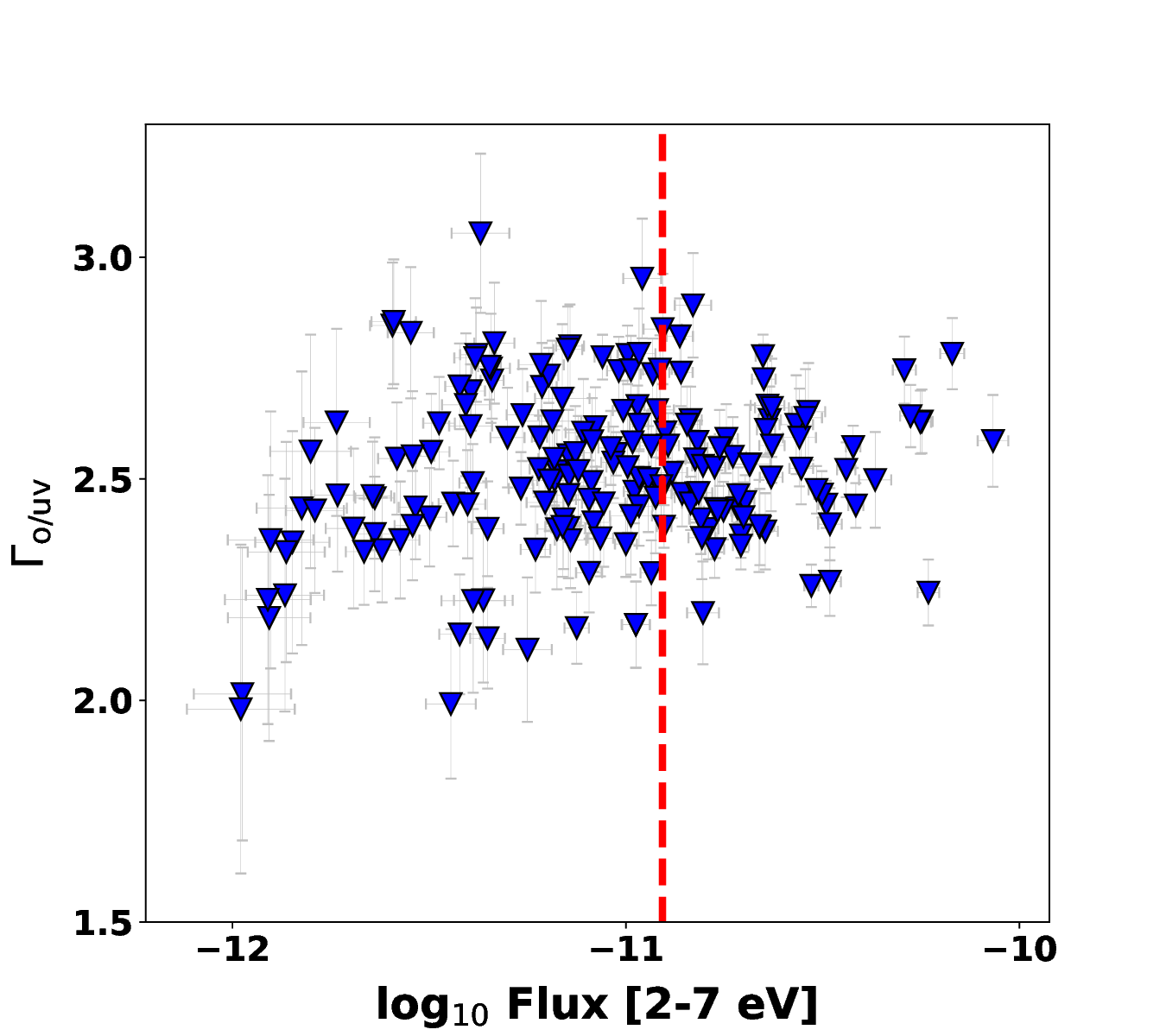}
		\caption{}
	\end{subfigure}
	\hfill
	\begin{subfigure}{0.48\textwidth}
		\includegraphics[width=0.85\textwidth]{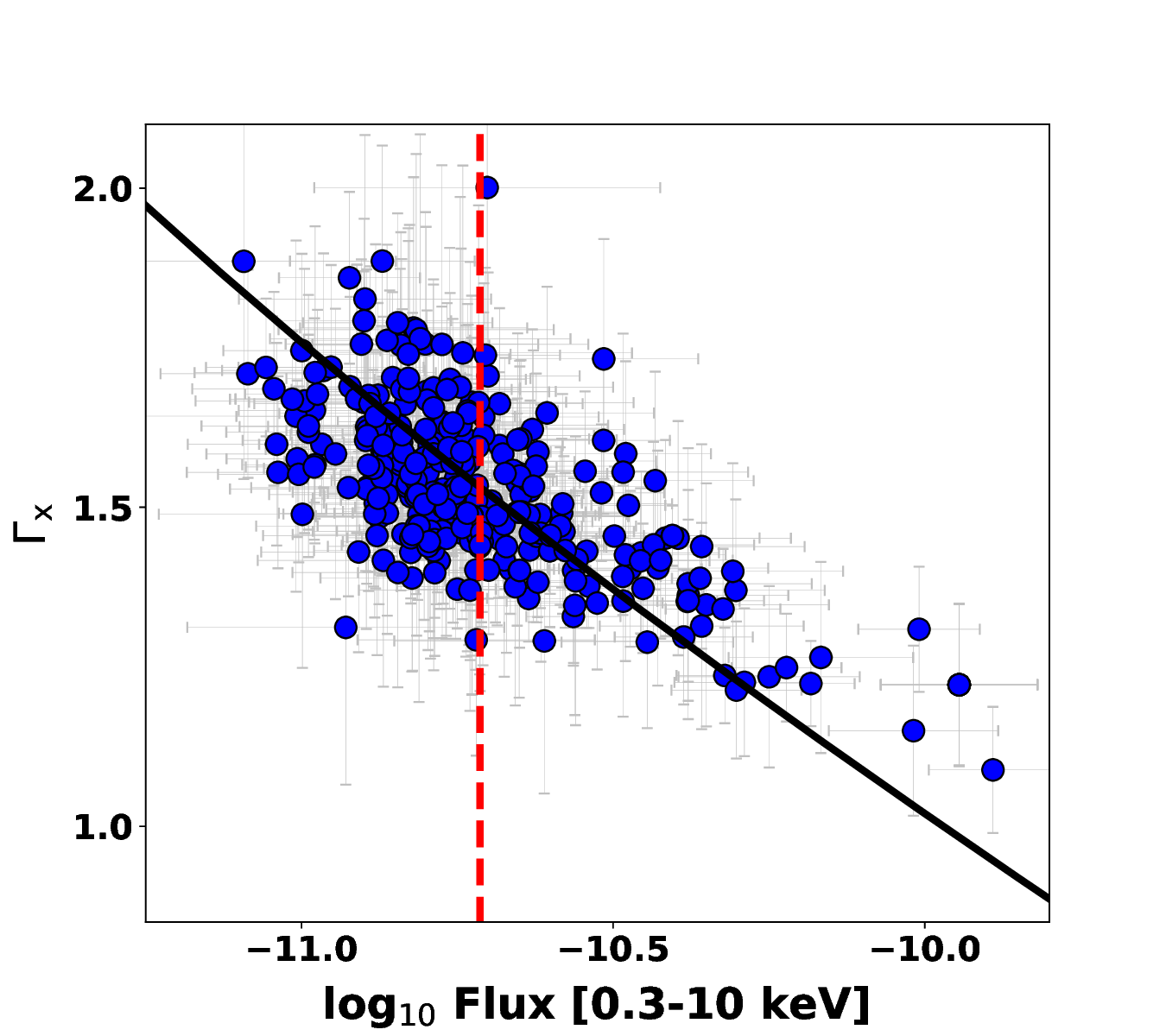}
		\caption{}
	\end{subfigure}
	
	\begin{subfigure}{0.49\textwidth}
		\includegraphics[width=0.8\textwidth]{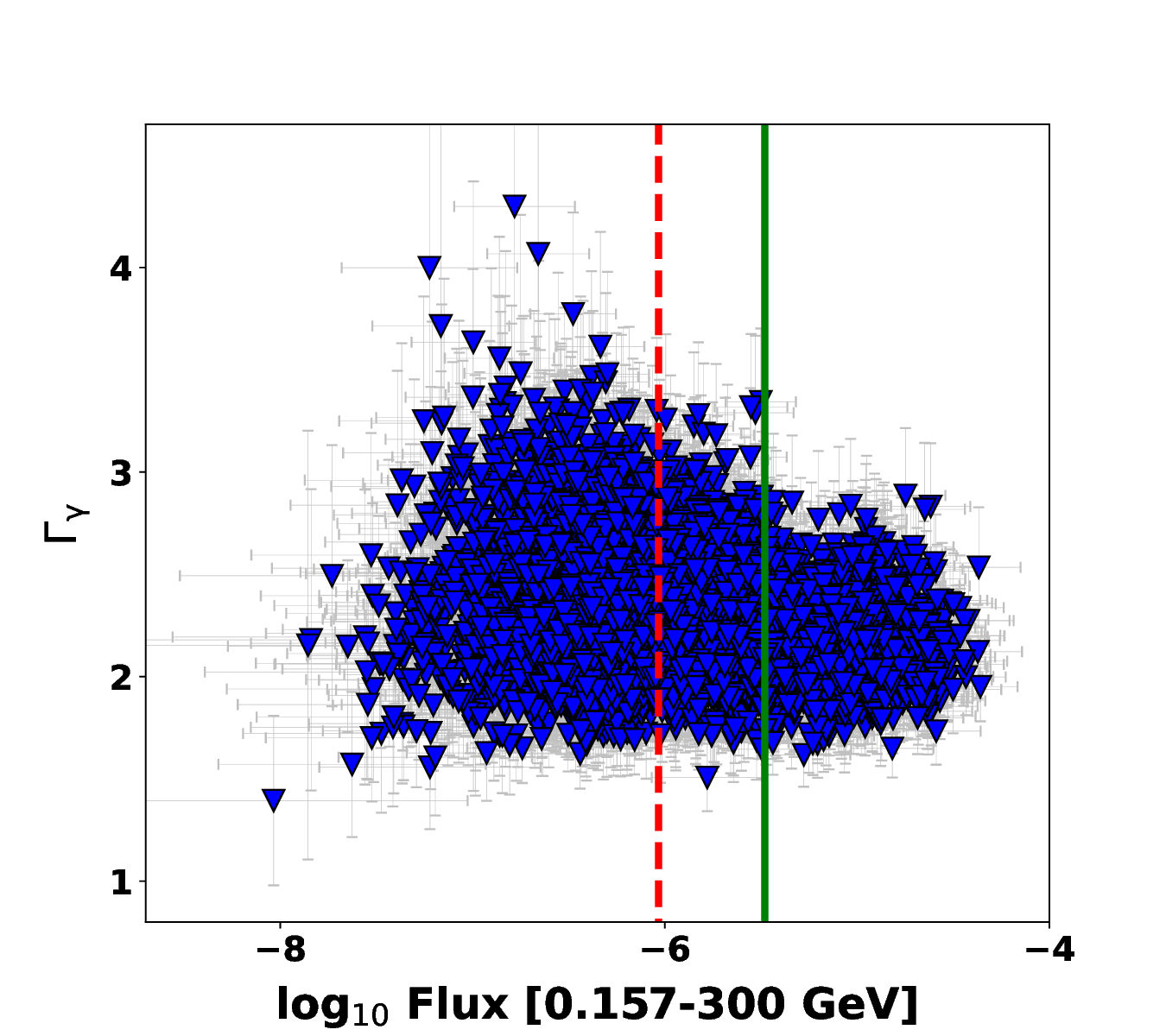}
		\caption{}
	\end{subfigure}
	
	\caption{Scatter Plots between Integrated fluxes and photon indices obtained through power-law fit in optical/UV , X-ray and $\gamma$-ray energies. 
	The vertical dashed lines represents the average value of corresponding fluxes in logscale. The X-ray and optical/UV fluxes are in units 
	of \textbf{ergs}\,\textbf{cm}$^{-2}$\,\textbf{s}$^{-1}$ and $\gamma$-ray fluxes are in units of \textbf{phs}\,\textbf{cm}$^{-2}$\,\textbf{s}$^{-1}$ 
	(a): The scatter plot between optical/UV flux and index. (b): The plot showing the anticorrelation of X-ray flux and index. the solid curve represents 
	the integrated power-law function mentioned in \S\ref{corre1}. (c): Scatter plot between integrated flux (0.157-300 GeV) and index of $\gamma$-ray spectrum (adaptive binning). 
	The dashed dotted vertical line represents the value of flux corresponding to the point of intersection of two Gaussians observed in the $\gamma$-ray flux distribution (\S\ref{distribution}).}
	\label{fig:index-flux}
\end{figure*}

\paragraph*{}
In Figures \ref{fig:index-flux} (a), (b), and (c), we show the scatter plots between the integrated fluxes at optical/UV, X-ray, and
$\gamma$-ray energies with their corresponding spectral indices. The dashed vertical line demarcates the 
average flux. A negative correlation between the flux
and indices indicates a ``harder when brighter'' trend while the reverse indicates a ``softer when brighter'' 
trend. Since we have considered the observations spanning over a decade, this study will tell the general 
behaviour of the source irrespective of its individual flaring states. It is evident from Figure \ref{fig:index-flux} (a) and \ref{fig:index-flux} (c), that no 
significant correlation was observed between the optical flux versus the index and the $\gamma$-ray flux versus
the index. A Spearman rank correlation study between these quantities resulted in the
correlation coefficient $\rho \approx 0.13$ with the null hypothesis probability $P\approx 0.06$ for the former, and $\rho =-0.09$ with
$P\approx 0.28$ for the latter. However, a negative correlation with the $\rho = -0.64$ and $p<0.001$ is obtained for 
the X-ray flux and index study. This suggests that the source
generally shows a ``harder when brighter'' trend in X-ray. A ``harder when brighter'' trend may be evident in time periods around the flares in optical/UV and $\gamma$-ray energies. However the longterm spectral analysis
indicates that this trend may not be the only possibility in these energy bands.

The negative correlation observed at X-ray energy also suggests that the enhancement in integrated 
flux may be predominantly due to the spectral hardening of the power-law function. 
The integrated flux in $\nu f_{\nu}$ representation between the photon energies $\epsilon_1$ and $\epsilon_2$ 
will be:
\begin{equation}
	\left(\frac{K\,\epsilon_{0}^{\rm \Gamma}}{2-\Gamma}\right) \left(\epsilon_{2}^{\rm 2-\Gamma}-\epsilon_{1}^{\rm 2-\Gamma}\right)
\end{equation}
Here, $K$ is the normalization, $\epsilon_0$ is the pivot energy and $\Gamma$ is the spectral index. 
To verify our inference, we plotted the integrated flux
against $\Gamma$ for a constant $K$. With a proper choice of $K$, we find that the predicted line 
reasonably supports the `harder when brighter' trend. However, the spread of the 
points around this line suggests, the variation in normalization is also
responsible for the flux variation. The absence of any such correlation at optical/UV and $\gamma$-ray
energies suggests significant variation in normalization may be 
responsible for the flux variations. Alternatively, this could also suggest that the source exhibits both 
``harder when brighter'' and ``softer when brighter'' behaviours during different
flaring states.
\subsubsection{Index-index Correlations}
\label{index-corre}
\begin{figure*}
	\begin{subfigure}{0.47\textwidth}
		\includegraphics[width=0.85\textwidth]{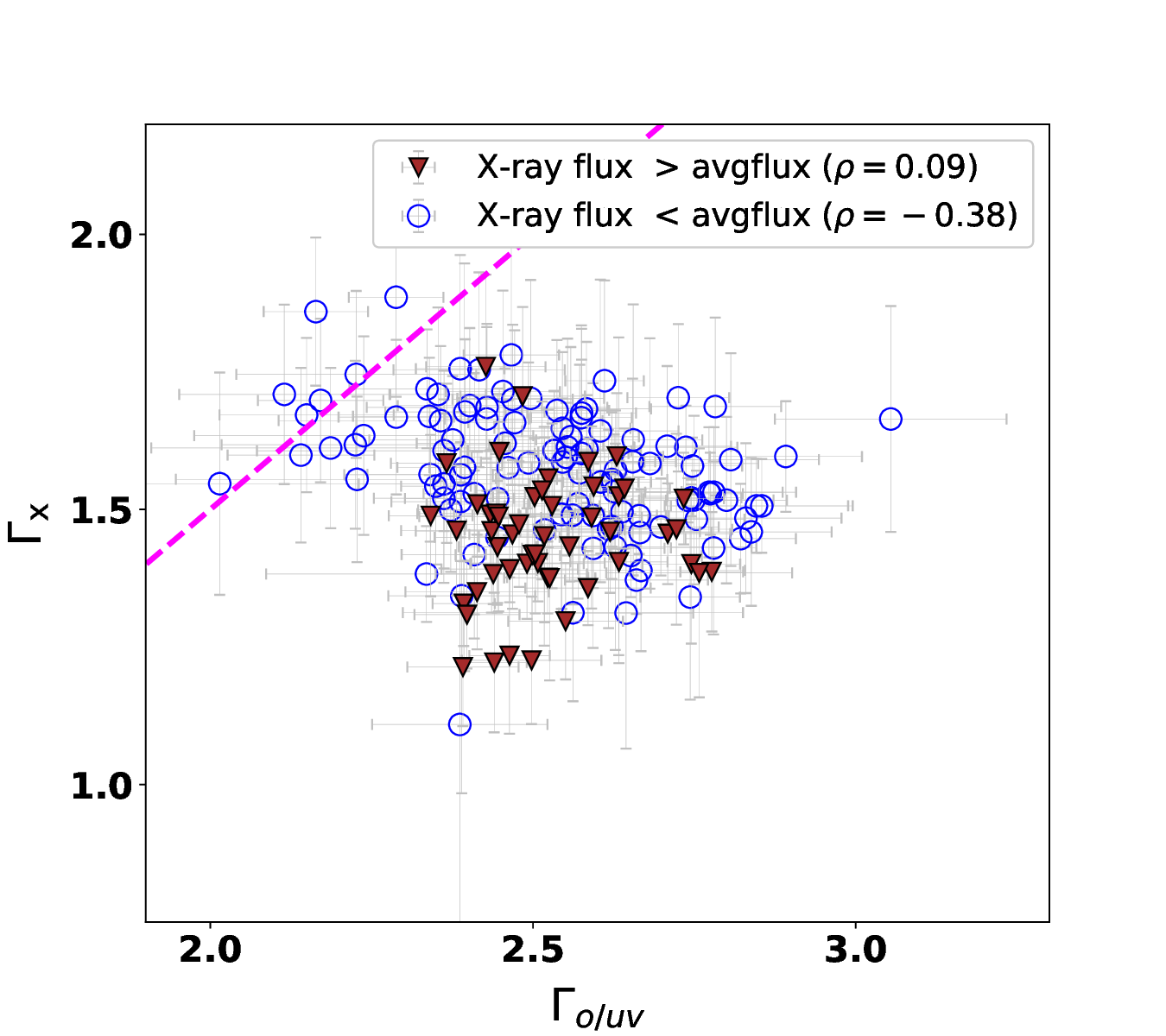}
		\caption{}
	\end{subfigure}
	\hfill
	\begin{subfigure}{0.47\textwidth}
		\includegraphics[width=0.85\textwidth]{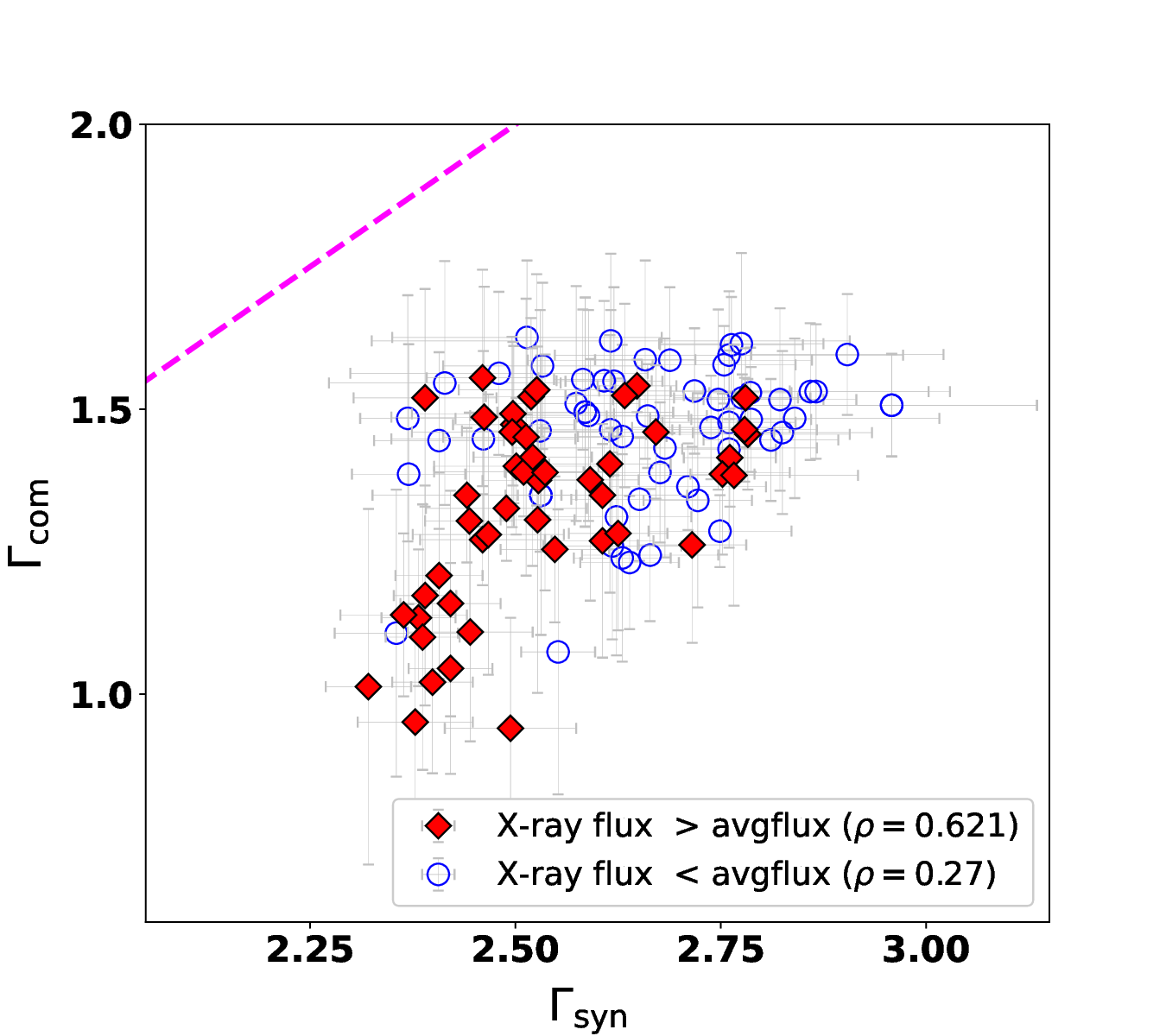}
		\caption{}
	\end{subfigure}
	
	\smallskip
	\begin{subfigure}{0.47\textwidth}
		\includegraphics[width=0.85\textwidth]{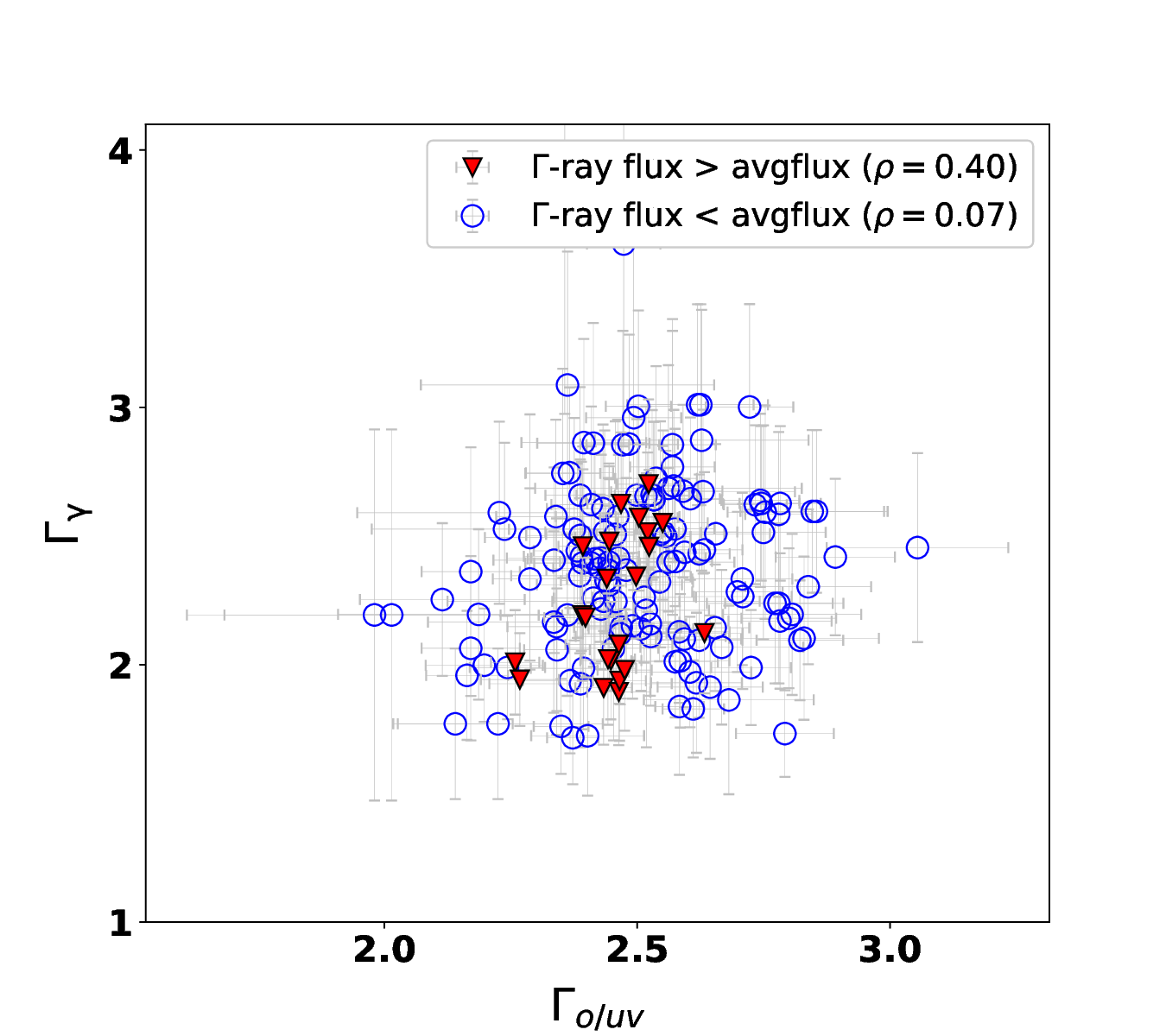}
		\caption{}
	\end{subfigure}
	\hfill
	\begin{subfigure}{0.47\textwidth}
		\includegraphics[width=0.85\textwidth]{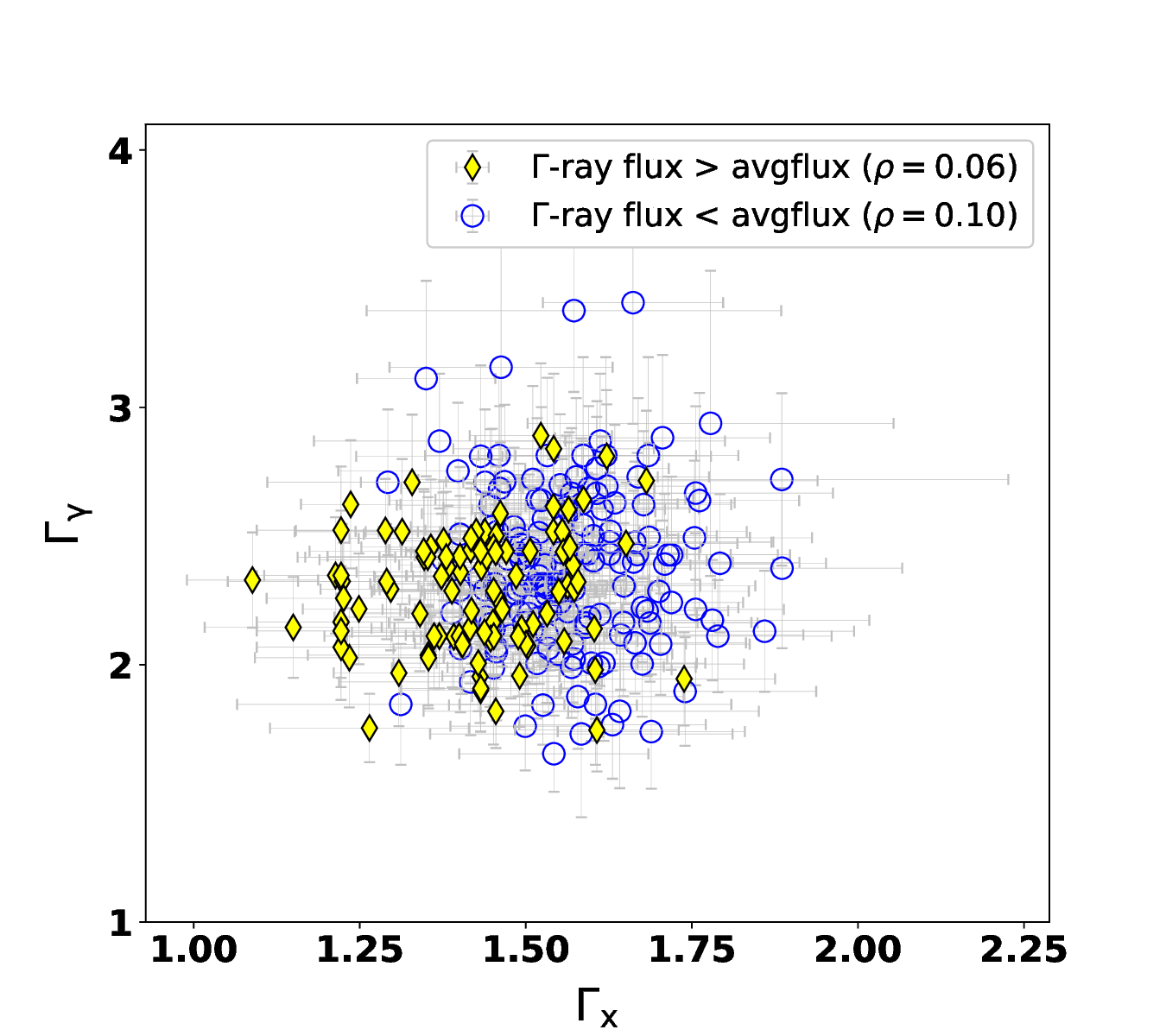}
		\caption{}
	\end{subfigure}
	
	\caption{Scatter plot between the spectral indices obtained through power-law and double power-law spectral fitting to simultaneous optical/UV, X-ray and $\gamma$-ray spectra. (a): Plot between optical/UV and X-ray spectral indices obtained through simple power-law fit ($\Gamma_{\rm o/uv}$ \& $\Gamma_{x}$). (b): Plot between optical/UV ($\Gamma_{\rm syn}$) and X-ray ($\Gamma_{\rm com}$) spectral indices obtained from double power-law fit. In the plots (a) \& (b), the dashed line indicate a straightline of equation `X\,=\,Y+0.5'. It is evident from the figures, that the points deviate considerably from this line indicating an index difference > 0.5 as mentioned in section \S \ref{index-corre}. (c): The scatter plot between optical/UV and $\gamma$-ray power-law indices. (d): Scatter plot between X-ray and $\gamma$-ray power-law indices. In all the subplots, filled shapes represent the states of higher flux and open shapes represent the low flux states and the details of correlation coefficient $\rho$  with respect to flux states have been inboxed in the plots.}
	
	\label{index-fig}	
\end{figure*}

\paragraph*{}
If we consider the radiative loss origin of the broken power-law electron 
distribution, then the indices should be positively correlated \citep{1962SvA.....6..317K, 2022MNRAS.514.3074B}. 
Since, $\Gamma_{\rm o/uv} > 2$ and $\Gamma_x < 2$ (Tables \ref{tab:xrt} \& \ref{tab:uvot}), these indices can relate to the high energy and the low 
energy power-law indices of the electron distribution. However, the Spearman correlation
study between these two quantities obtained from the power-law fit suggested a poor correlation 
with $\rho\approx -0.23$ and $P\approx 0.002$. The scatter plot between $\Gamma_{\rm o/uv}$ and  $\Gamma_x$
is shown in Figure \ref{index-fig} (a). Interestingly, a moderate positive correlation with $\rho \approx 0.52$
and $P < 0.001$ is witnessed (figure \ref{index-fig} (b)) when the optical/UV and X-ray spectral indices are obtained from 
a double power-law function ($\Gamma_{\rm syn}$ and  $\Gamma_{\rm com}$). 
However, the difference between the optical/UV 
and X-ray spectral indices are much larger than 0.5 and hence, cannot be interpreted in 
terms of radiative losses \citep{2022MNRAS.514.3074B}. Therefore, this study is inconclusive regarding the radiative 
cooling interpretation of the broken power-law electron distribution. 

If the electron distribution responsible for the broadband emission is an outcome of multiple 
acceleration processes (e.g. stochastic and shock acceleration), then the resultant
broken power-law indices will be governed by the corresponding acceleration rate \citep{2008MNRAS.388L..49S}.
Under this scenario, it may happen that the power-law indices of the electron distribution (or the
corresponding photon spectral indices) may not be strongly correlated. Such a study may involve the 
exact description of the acceleration processes with a suitable choice of the magneto-hydrodynamic 
turbulence \citep{2007Ap&SS.309..119R, 2019Galax...7...78R}. This will involve additional parameters which 
may not be constrained and also beyond the scope of the present work. 

We also did not find any 
significant correlation between $\Gamma_{\rm o/uv}$ and $\Gamma_x$ with the $\gamma$-ray 
spectral indices $\Gamma_\gamma$. This result may indicate that the electron energies responsible
for the emission at these photon energies may be different. Probably, the electron population
responsible for the $\gamma$-ray emission may fall close to the break energy. This is consistent
with the range of $\Gamma_\gamma$ which is spread around 2.

To study the spectral index behaviour of the source during high and low flux states we demarcate
the points in the scatter plots in Figure \ref{index-fig} with filled (high flux) and open (low flux)
shapes. From the scatter plot between the indices obtained from power-law spectral fit of 
optical/UV and X-ray spectra, we find during high flux states (filled shapes) the $\Gamma_x$ is 
moderately harder and $\Gamma_{\rm o/uv}$ softer. However, no such behaviour is seen when the 
indices are obtained through a double power-law function (Figure \ref{index-fig} (b)). Similarly, no definite behaviour
was observed during the high/low flux states of the $\gamma$-ray spectrum. In Figure \ref{index-fig} (c) and (d), we divide 
the data points on the basis of average $\gamma$-ray flux with the filled/open shapes indicating the
high/low $\gamma$-ray flux. The scatter of the filled data points are almost similar to the open ones
with $\Gamma_\gamma$ spread around 2. 
The individual flares may exhibit
either ``bluer when brighter'' or ``redder when brighter'' which is consistent with the no definite
pattern in the scatter plot. On the other hand, the X-ray indices during the $\gamma$-ray high flux show a 
mild hardness (figure \ref{index-fig} (c))which is consistent with the ``harder when brighter'' trend mentioned earlier.
\subsubsection{Flux-Flux Correlations}
\label{flux-corre}
\begin{figure*}
	\begin{subfigure}{0.47\textwidth}
		\includegraphics[scale=0.7]{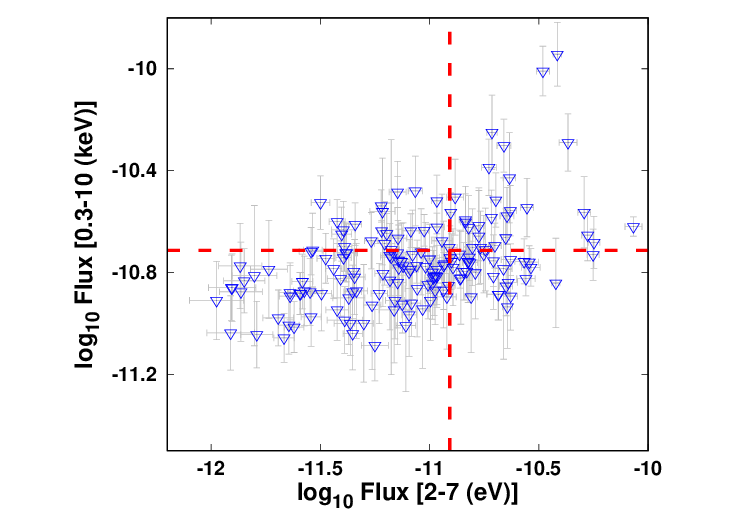}
		\caption{}
	\end{subfigure}
	\hfill
	\begin{subfigure}{0.47\textwidth}
		\includegraphics[scale=0.7]{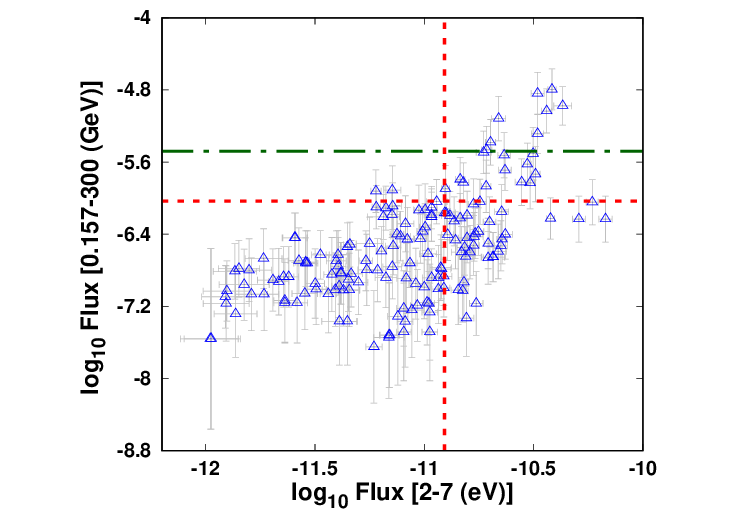}
		\caption{}
	\end{subfigure}
	
	\begin{subfigure}{0.47\textwidth}
		\includegraphics[scale=0.7]{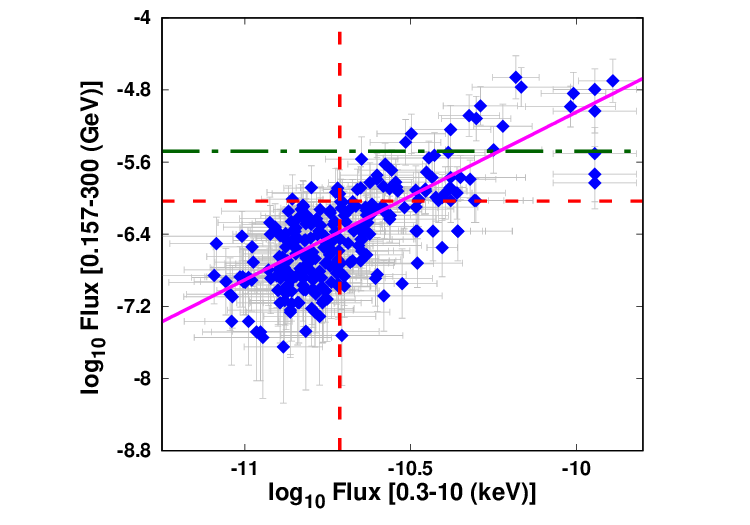}
		\caption{}
	\end{subfigure}
	\caption{Scatter plot between the fluxes of optical/UV , X-ray and $\gamma$-ray spectra of 3C\,279. (a): Plot between X-ray and optical/UV flux. (b): Plot between optical/UV and $\gamma$-ray flux.
	(c): Plot showing strong correlation between X-ray and $\gamma$-ray flux. The solid line represents the best fitted linear regression to this correlation. The dashed lines in all the plots represents the value of arithmetic average of the fluxes plotted on respective axis. The X-ray and optical/UV fluxes are in units of \textbf{ergs}\,\textbf{cm}$^{-2}$\,\textbf{s}$^{-1}$ and $\gamma$-ray fluxes are in units of \textbf{phs}\,\textbf{cm}$^{-2}$\,\textbf{s}$^{-1}$. The dotted dashed horizontal line in (b) and (c) represent the value of flux corresponding to the point of intersection of two Gaussians observed in the $\gamma$-ray flux distribution (\S\ref{distribution})}
	\label{flux-fig}
\end{figure*}

\paragraph*{}
3C\,279 was known to exhibit multi-wavelength flux correlations during its several major outbursts 
\citep{2015ApJ...807...79H, 2021arXiv210311149Z, 2012ApJ...754..114H, 2018MNRAS.479.2037P, 2020ApJ...890..164P, 2015ApJ...808L..48P, 2020MNRAS.498.5128R}.
To study whether such correlations exist in its long term behaviour spanning over $\sim$ 14 years, we performed
a correlation test on the simultaneous fluxes at the optical/UV, X-ray, and $\gamma$-ray energies.
Contrary to X-ray--$\gamma$-ray spectral indices, the $\gamma$-ray flux is well correlated with the X-ray flux,
and the plot between these quantities is shown in Figure \ref{flux-fig} (c). The Spearman rank correlation study 
resulted in $\rho$\,=\,0.72 with $P<0.001$. 

The flux--index correlation study at X-ray energies suggests the flux variability is predominantly due to the
change in the index. Wheraas at $\gamma$-ray energy no such correlation is observed. However, the range of index values suggests that 
the Compton SED peak may fall at this energy range. Hence, flux--index correlation
at X-rays can also cause variation in flux at the low energy $\gamma$-rays if the same power-law electron
distribution is responsible for the emission at both these energies.

The observed correlation between the X-ray and $\gamma$-ray fluxes is also possible when the
same radiative process is responsible for the emission at these energies. 
To understand this, we performed a 
linear regression analysis between the logarithm of the X-ray and $\gamma$-ray fluxes and obtained the best fit straight line as 
\begin{align}
\log F_{\rm 0.157-300 GeV}=(1.86\pm 0.1)\log_{10} F_{\rm 0.3-10 keV}+(13.55\pm 1.0)
\end{align}
with Q-value of $0.87$. A near quadratic dependence of the $\gamma$-ray flux on the X-ray flux disfavours this
interpretation. Quadratic dependence between the X-ray and $\gamma$-ray fluxes is also observed
for the BL Lac type source MKN\,421 \citep{2019MNRAS.484.2944G, 2018ApJ...854...66K, 1997A&A...320...19M, 2020MNRAS.499.2094G}. 
The X-ray emission for this source is due to the synchrotron process
and the quadratic dependence of $\gamma$-ray emission supports the SSC interpretation of the Compton 
spectral component \citep{2005ApJ...630..130B, 2011ApJ...738...25A}.
However, in the case of 3C\,279, the X-ray and the $\gamma$-ray emission are both due to IC process.
The broadband SED modelling of the source also associates the X-ray emission to the SSC process and the 
$\gamma$-ray emission to the EC process (\S\ref{broadband}).

A moderate correlation is also observed between the optical/UV and $\gamma$-ray fluxes with the Spearman 
rank correlation study resulting in $\rho=0.60$ and $P<0.001$. Again, this correlation may be associated with the fact that 
the Compton peak falls in $\gamma$-ray regime. If we consider a broken power-law electron
distribution responsible for the broadband emission, then the enhancement 
in optical flux may be associated with the increase in high energy
(greater than the break energy) electron component. This would also affect the high energy $\gamma$-ray
flux resulting in the observed correlation. However, the correlation may be suppressed due to the orphan
flares often encountered in 3C\,279 \citep{2020MNRAS.498.5128R, 2021ApJS..257...37P}.
We did not observe any significant correlation between optical/UV and the X-ray fluxes. The Spearman 
correlation study resulted in $\rho =0.42$ and $P<0.001$. This is consistent with the absence of a strong
correlation between the corresponding spectral indices.
\subsubsection{Correlation with Transition Energy}
\label{trans-corre}
\paragraph*{}
We identify the transition energy as the photon energy at which the dominance of the 
synchrotron or IC flux switches. This is obtained by fitting the optical/UV and the X-ray 
spectra with broken power-law and double power-law functions (\S\ref{trans}). Consistently, we obtained
two estimates of the transition energy, $\epsilon_b$ and $\epsilon_v$, and studied the correlation
of these quantities with the spectral indices and fluxes. Since the transition energy is very sensitive
to the indices, even a minor change in the latter will vary the result considerably. We find both $\epsilon_b$ and $\epsilon_v$
are well correlated with the optical/UV flux though, the correlation was much stronger in the case of
$\epsilon_v$. The correlation results are $\rho = 0.73$ with $P<0.001$ for $\epsilon_b$ and 
$\rho = 0.91$ with $P<0.001$ for $\epsilon_v$. We find no significant correlation 
of the transition energy with the optical/UV index and the correlation results are $\rho = -0.33$ with 
$P<0.001$ in the case of $\epsilon_b$ and $\Gamma_{\rm o/uv}$ while $\rho = -0.39$ with $P<0.001$ for $\epsilon_v$
and $\Gamma_{\rm syn}$. 

\begin{figure*}
	\begin{subfigure}{0.47\textwidth}
		\includegraphics[width=0.86\textwidth]{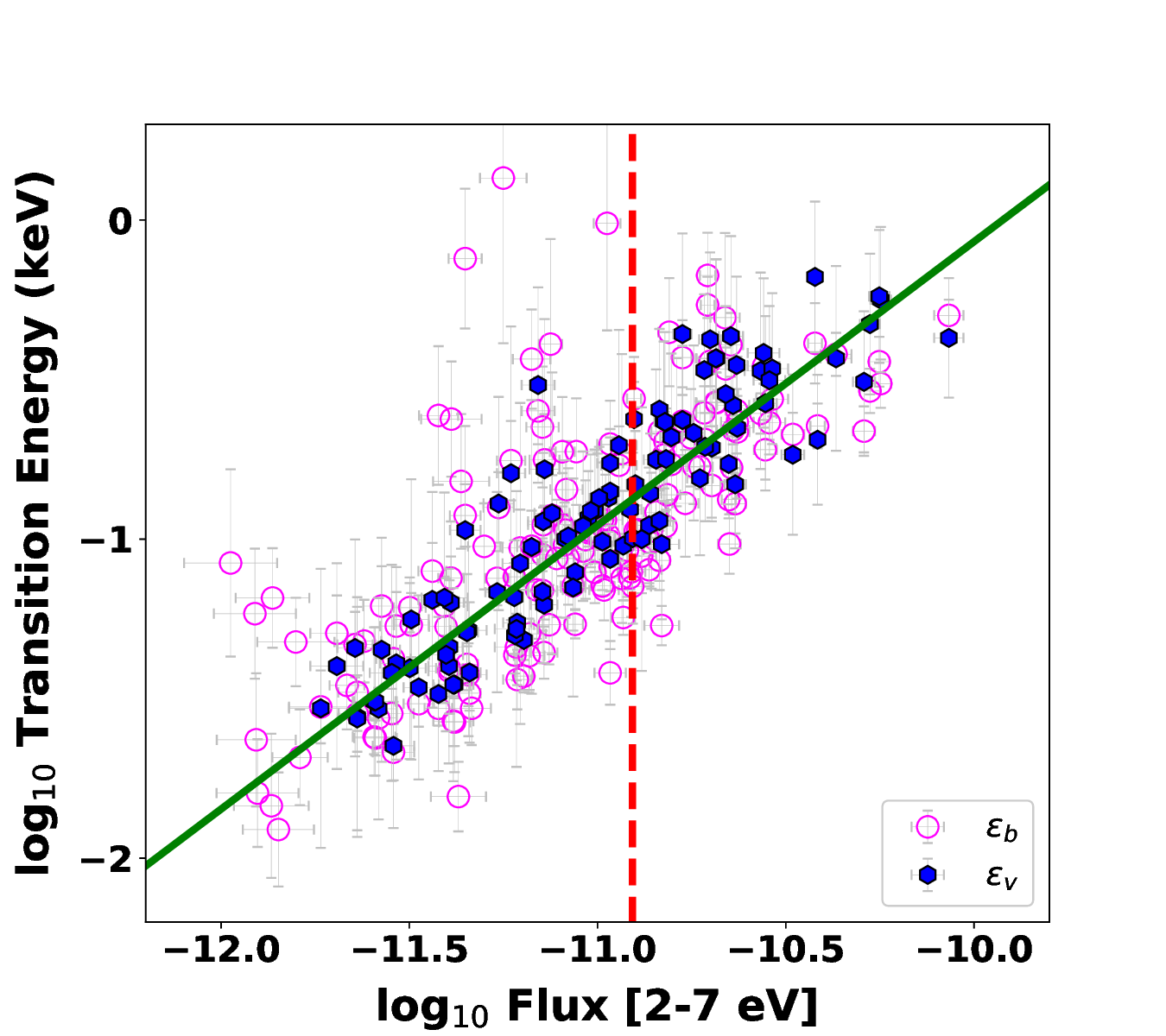}
		\caption{}
	\end{subfigure}
	\hfill
	\begin{subfigure}{0.47\textwidth}
		\includegraphics[width=0.86\textwidth]{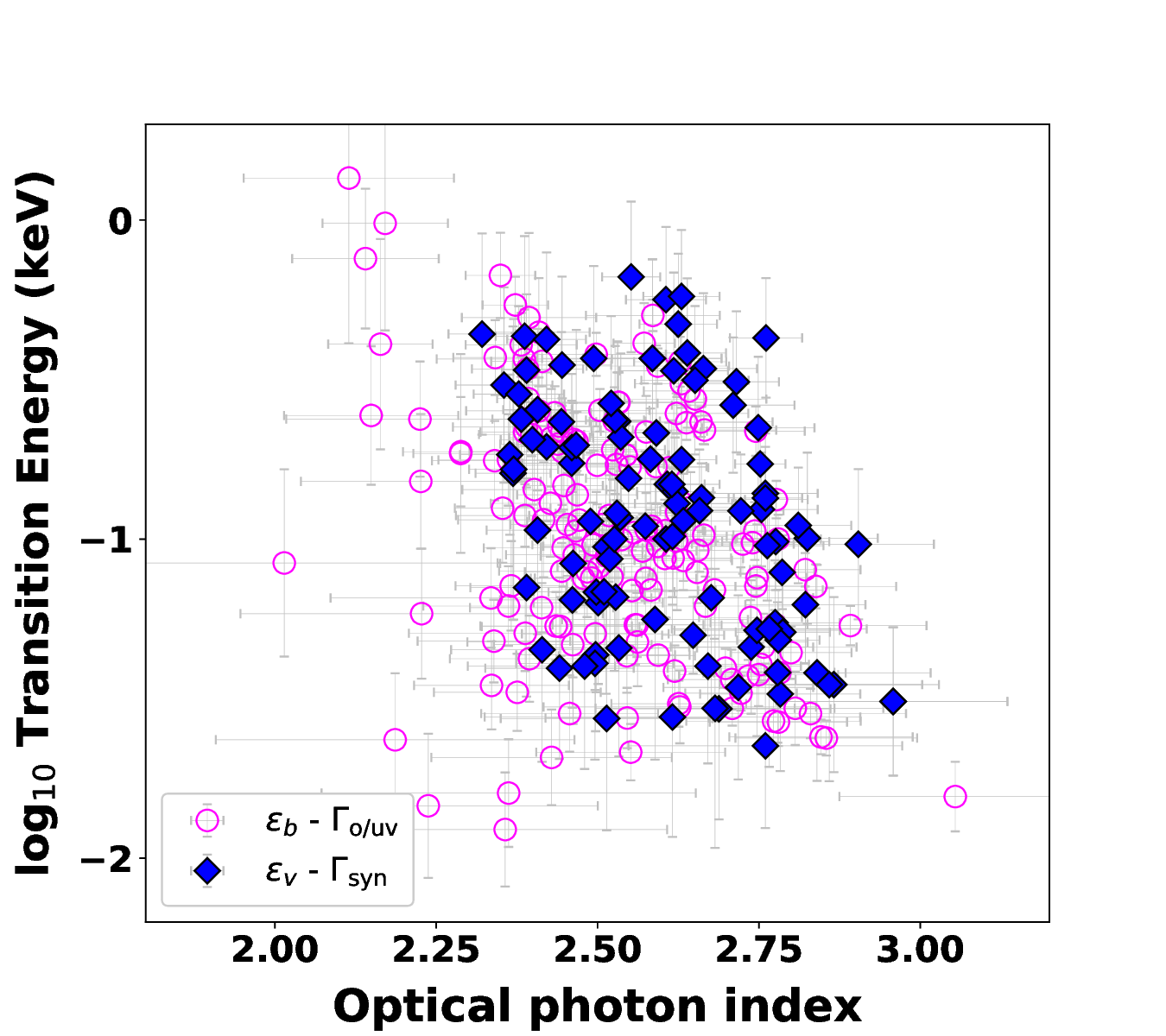}
		\caption{}
	\end{subfigure}
	
	\smallskip
	\begin{subfigure}{0.47\textwidth}
		\includegraphics[width=0.86\textwidth]{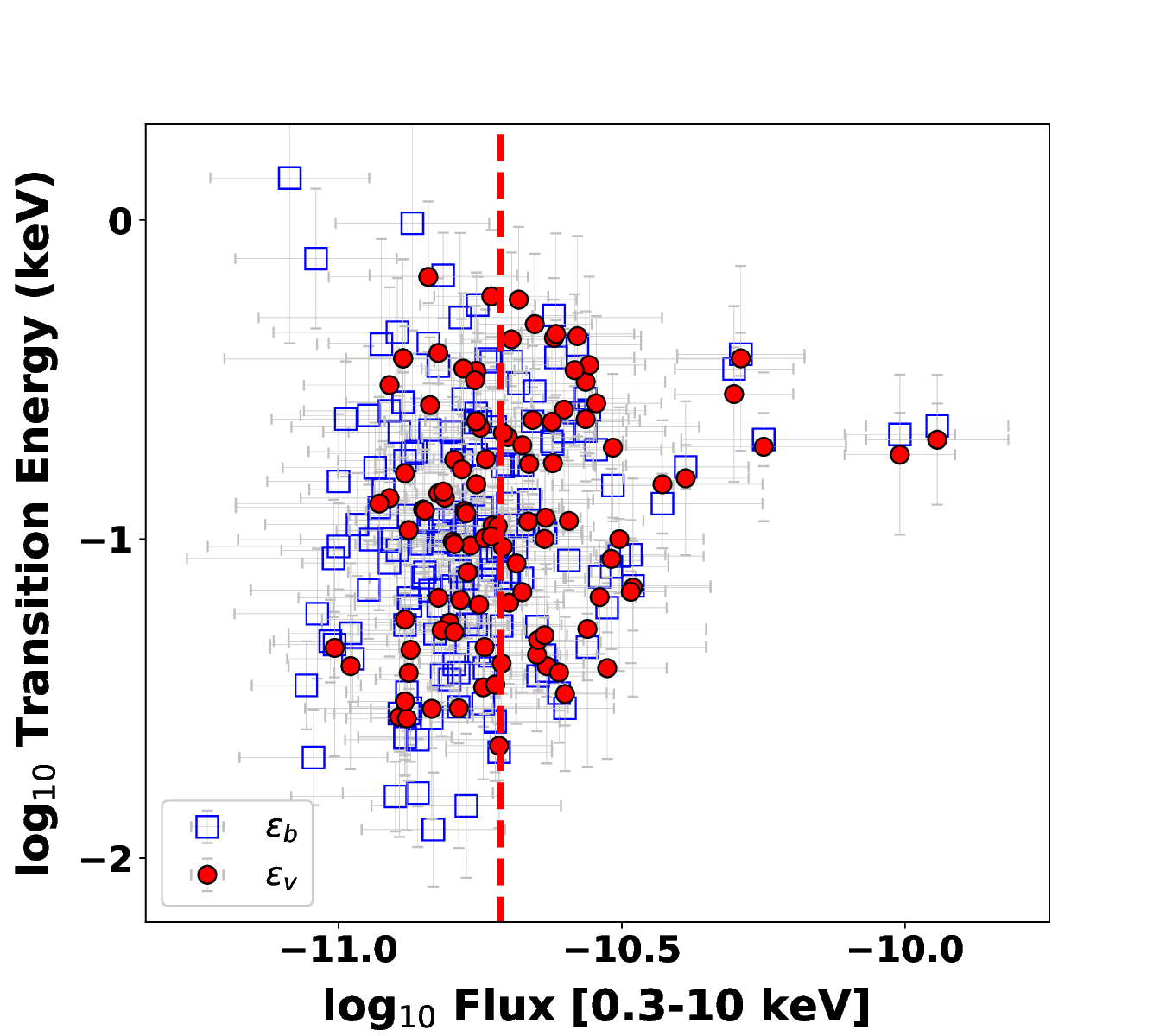}
		\caption{}
	\end{subfigure}
	\hfill
	\begin{subfigure}{0.47\textwidth}
		\includegraphics[width=0.86\textwidth]{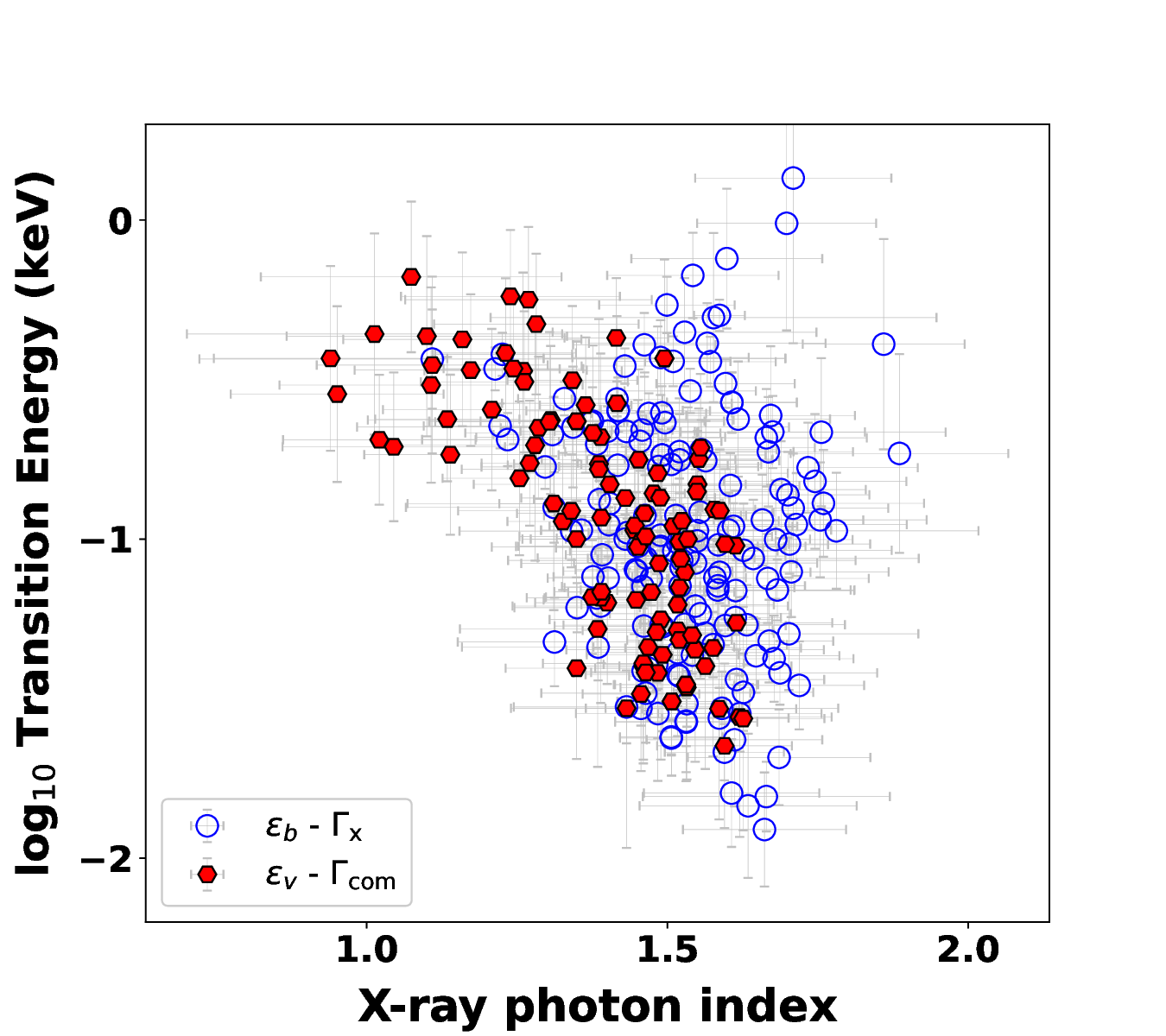}
		\caption{}
	\end{subfigure}
	
	\caption{Scatter plot between transition energy and other parameters. The X-ray and optical/UV fluxes are in units of \textbf{ergs}\,\textbf{cm}$^{-2}$\,\textbf{s}$^{-1}$. (a): Hollow circles represents the variation of $\epsilon_b$ with optical/UV flux and filled ones show the variation of $\epsilon_v$ with optical/UV flux. The solid line gives the best fit linear regression to $\epsilon_v$-flux correlation. The vertical dashed line represents the average optical/UV flux. (b): Hollow circles represents the variation of $\epsilon_b$ with $\Gamma_{\rm o/uv}$ and filled diamond shapes show the variation of $\epsilon_v$ with $\Gamma_{\rm syn}$. (c): Hollow squares represent the variation of $\epsilon_b$ with X-ray flux and filled circles show the variation of $\epsilon_v$ with X-ray flux. The vertical dashed line gives the average X-ray flux in logscale. (d): Hollow circles represents the variation of $\epsilon_b$ with $\Gamma_{x}$ and filled points show the strong negative correlation of  $\epsilon_v$ with $\Gamma_{\rm com}$.}
	\label{fig:ev}
\end{figure*}

The correlation results, in the case of transition energy with X-ray spectrum were contrary to that of
optical/UV. We found that the transition energy is anti-correlated with the X-ray spectral index 
with $\rho = -0.72$ and $P<0.001$ in the case of $\epsilon_v$ and $\Gamma_{\rm com}$. However, this correlation is
found to be weak in the case of $\epsilon_b$ and $\Gamma_x$ where we obtained 
$\rho = -0.14$ and $P=0.07$. This is consistent with the variation observed between $\Gamma_x$ and $\Gamma_{\rm com}$ (figure \ref{fig:eb-ev}(c)). 
No significant correlation was observed between the transition energy and the X-ray flux. The correlation results are $\rho=0.15$ with $P=0.04$ 
in the case of $\epsilon_b$ while, $\rho=0.20$ with $P=0.03$ for $\epsilon_v$ with X-ray flux.
The scatter plot between the transition energy and the flux/index is shown in Figure \ref{fig:ev} with the vertical 
dashed lines in (a) and (c) denoting the average optical/UV and X-ray fluxes.

To understand the variability behaviour of the source depicted by these correlation studies with the transition energy, we performed
a linear regression analysis between the quantities which showed a strong correlation. We were able to 
obtain a reasonable fit in the case of $\epsilon_v$ and $F _{\rm 2-7 eV}$ and the linear relation obtained is given by
\begin{align}
	\log(\epsilon_v)\,=\,(0.89\pm 0.04)\,\log(F_{\rm 2-7\, eV})\,+\,(8.82\pm 0.44)
\end{align}
with Q-value of 0.99.
The near-linear dependence of the transition energy with the optical/UV flux supports that 
the flux variability at this energy band is mainly governed by the changes in the normalization.
This inference is further asserted by the absence of any correlation between the transition
energy and the optical/UV spectral index. The anti-correlation between the transition energy and
the X-ray spectral index is consistent with the ``harder when brighter'' trend observed in the 
X-ray flux-index correlation. No appreciable correlation between the transition energy and the X-ray 
flux also supports that the flux variations are mainly governed by the index changes in this band. If the broken
power-law electron distribution responsible for the broadband emission is given by
\begin{align}
	N(E)\,dE = 
	\begin{cases}
		K\, E^{-p_1}dE \quad &\textrm{for} \; E\,\leq E_b\\
		K\, E_b^{p_2-p_1} E^{-p_2} dE\quad  &\textrm{for} \; E \ge E_b
	\end{cases}
\end{align}
then extending our correlation/regression studies to the emitting electron distribution, one can 
conclude that: {\it at low energy ($E<E_b$) the variations are mainly governed by the changes in index $p_1$,
whereas at high energy ($E>E_b$) the variations can be due to the change in $E_b$}. Since the X-ray spectrum is
governed by the low energy electrons, the changes in $p_1$ are consistent with the ``harder when brighter''
trend observed in the flux-index correlation and the anti-correlation observed between the transition energy 
and X-ray spectral index. Similarly, the transition energy-flux correlation observed for the optical/UV band
is consistent with the change in normalization initiated by the variations in $E_b$. The change in $E_b$
may also be consistent with the variation in $\gamma$-ray peak suggested by the range of $\Gamma_\gamma$.
These results disfavour the radiative cooling interpretation of $E_b$ and probably the broken power-law
electron distribution may be an outcome of the acceleration process itself \citep{2003A&A...410..813A, 2008MNRAS.388L..49S, 2007Ap&SS.309..119R}.
The details of all Spearman rank correlation studies discussed in this section are also funished in Table \ref{tab:corr}.
\begin{table*}
	\centering
	\tabcolsep 16.0pt
	\setlength\extrarowheight{1pt}
	\scriptsize
	\begin{tabular}{lrl} 
		\hline
		\hline
		Parameters/quantities & $\rho \ \ $ & $P$ \\
		\hline
		\hline	
		
		$\Gamma_{\rm o/uv}$ and Flux [2-7 (eV)] & 0.13 &  0.06\\
		$\Gamma_{x}$ and Flux [0.3-10 (keV)] & -0.64 & <0.001\\
		$\Gamma_{\rm \gamma}$ and Flux [0.157-300 (GeV)] & -0.09 & 0.28\\
		$\Gamma_{\rm o/uv}$ and $\Gamma_{x}$ & -0.23 &  0.002\\
		$\Gamma_{\rm o/uv}$ and $\Gamma_{\rm \gamma}$  & -0.03 & 0.68 \\
		$\Gamma_{x}$ and $\Gamma_{\rm \gamma}$  & 0.07 & 0.38\\
		Flux [2-7 (eV)] and Flux [0.3-10 (keV)] & 0.42 & <0.001 \\
		Flux [2-7 (eV)] and Flux [0.157-300 (GeV)] & 0.60 & <0.001\\
		Flux [0.3-10 (keV)] and Flux [0.157-300 (GeV)] & 0.72 & <0.001\\
		$\epsilon_b$  and Flux [2-7 (eV)] &  0.73 & <0.001\\
		$\epsilon_b$  and Flux [0.3-10 (keV)] &  0.15 & 0.04\\
		$\epsilon_b$ and $\Gamma_{\rm o/uv}$ & -0.33 & <0.001 \\
		$\epsilon_b$ and $\Gamma_{x}$ & -0.14 & $0.07$ \\
		
		$\epsilon_v$  and Flux [2-7 (eV)] &  0.91 & <0.001\\
		$\epsilon_v$  and Flux [0.3-10 (keV)] &  0.20 & 0.03\\
		$\epsilon_v$ and $\Gamma_{\rm syn}$ & -0.39 & <0.001 \\
		$\epsilon_v$ and $\Gamma_{\rm com}$ & -0.72 & <0.001 \\
		$\Gamma_{\rm syn}$ and $\Gamma_{\rm com}$ & 0.52 & <0.001 \\
		\hline
		$\Gamma_{\rm o/uv}$ and $\Gamma_{x}$ & 0.09 &  0.50\\
		(x-ray flux > average flux) & & \\
		
		$\Gamma_{\rm o/uv}$ and $\Gamma_{x}$ & -0.38 & <0.001\\ 
		(x-ray flux < average flux) & & \\
		$\Gamma_{\rm syn}$ and $\Gamma_{\rm com}$ & 0.62 & <0.001 \\
		(x-ray flux > average flux) & & \\
		$\Gamma_{\rm syn}$ and $\Gamma_{\rm com}$ & 0.27 & 0.12 \\
		(x-ray flux < average flux) & & \\
		
		$\Gamma_{\rm o/uv}$ and $\Gamma_{\rm \gamma}$  & 0.40 & 0.06 \\
		($\gamma$-ray flux > average flux) & & \\
		
		$\Gamma_{\rm o/uv}$ and $\Gamma_{\rm \gamma}$  & 0.07 & 0.38 \\
		($\gamma$-ray flux < average flux) & & \\
		
		$\Gamma_{x}$ and $\Gamma_{\rm \gamma}$  & 0.06 & 0.51\\
		($\gamma$-ray flux > average flux) & & \\
		
		$\Gamma_{x}$ and $\Gamma_{\rm \gamma}$  & 0.10 & 0.31\\
		($\gamma$-ray flux < average flux) & & \\
		
		\hline
		\hline
	\end{tabular}
	\caption{The details of correlation results found between different source parameters/quantities.}
	\label{tab:corr}
\end{table*} 	

\begin{table*}
\setlength\extrarowheight{3pt}
\scriptsize
\begin{tabular}{c c c c c}
	\hline
	\hline
	&  \multicolumn{1}{c}{Number of }   &  \multicolumn{1}{c}{Normal (Flux)}            &     \multicolumn{1}{c}{Normal (log Flux)} & \multicolumn{1}{c}{Normal (Spectral index)}\\ 
	& data points & AD(critical value) &  AD(critical value) &  AD(critical value) \\ \hline  
	
	Fermi-3 day binned  & 1354  & 219 (0.785) & 2.94 (>0.785) & 30.7 (0.785) \\
	\hline
	Fermi adaptive binned & 5850 & 916 (0.786) & 57.98 (0.786) & 31.80 (0.786)\\							   
	\hline
	
	X-ray & 326 & 39.20 (0.778) & 9.76 (0.778) & {\bf 0.65} (0.778)\\
	\hline
	
	optical/UV & 189 & 11.30 (0.771) & {\bf 0.45} (0.771) & {\bf 0.54} (0.771) \\
	\hline							   
	\hline
\end{tabular}
\caption{Results of the Anderson Darling test done for the flux/index distributions in $\gamma$-ray, X-ray and optical/UV bands of 3C\,279.}
\label{tab:ad}
\end{table*}
\section{Distribution of Fluxes and Indices}
\label{distribution}
\paragraph*{}
The FSRQ 3C\,279 exhibits prominent flares in various energy bands with significant spectral variations. 
There are epochs when these flares are correlated over all the energy bands and also for the peculiar orphan ones.  
For example, the GeV flares at epochs 56646, 56720, 56750, and 57188 have been previously reported as orphan $\gamma$-ray flares \citep{2017ApJ...850...87M,2019hepr.confE..75L,2022PhRvD.105b3005W,2021JHEAp..29...31P}. 
The spectral property of the source within orphan and multi-wavelength flares also differ considerably \citep{2020MNRAS.498.5128R}. 
Studying the long term flux distribution has the potential
to identify whether these flux variations are associated with a single statistical process or the source behaviour changes 
during different flux states \citep{2018RAA....18..141S}. Such studies on blazar lightcurve (including 3C\,279) 
at different energy bands have already been done and they suggest
a log-normal flux variability \citep{2012MNRAS.420..604N,2018RAA....18..141S,2003MNRAS.345.1271V,2018Galax...6..135R,2014ApJ...786..143S}. 
Since the thermal emission from the accretion disk is also log-normal in nature, this flux
variability of blazar was assumed to highlight the link between the accretion disk and the blazar jet \citep{2010LNP...794..203M,2012MNRAS.420..604N}.
However, such a flux distribution can also be an outcome of normal fluctuations in the power-law spectral index \citep{2018MNRAS.480L.116S,2020MNRAS.491.1934K}.
The $\gamma$-ray flux distribution of 3C\,279 shows a double-Gaussian nature suggesting two definite flux 
states \citep{2018RAA....18..141S}. We repeated this by including the optical/UV, X-ray, and $\gamma$-ray light curves, and also studied
the distribution of the spectral indices.

To study whether the flux and index variations are consistent with a normal distribution, we first performed the 
Anderson-Darling (AD) tests on these quantities. Depending on the value of the test statistics, one can reject the normality
of the distribution when it is greater than the critical value. The test was performed on the fluxes, logarithm of fluxes, and 
spectral indices. In Table \ref{tab:ad}, we provide the results of the AD test with the critical values estimated at 5\% significance 
of the null hypothesis. We find that the distributions of the logarithm of the optical/UV fluxes, optical/UV spectral indices 
and the X-ray spectral indices favour a Gaussian distribution (in bold); while the other distributions are inconclusive. 
This behaviour of the optical/UV flux is consistent with the earlier studies where the blazar flux variations are log-normal
in nature.

\begin{figure*}
	\begin{subfigure}{0.49\textwidth}
		\includegraphics[scale=0.74]{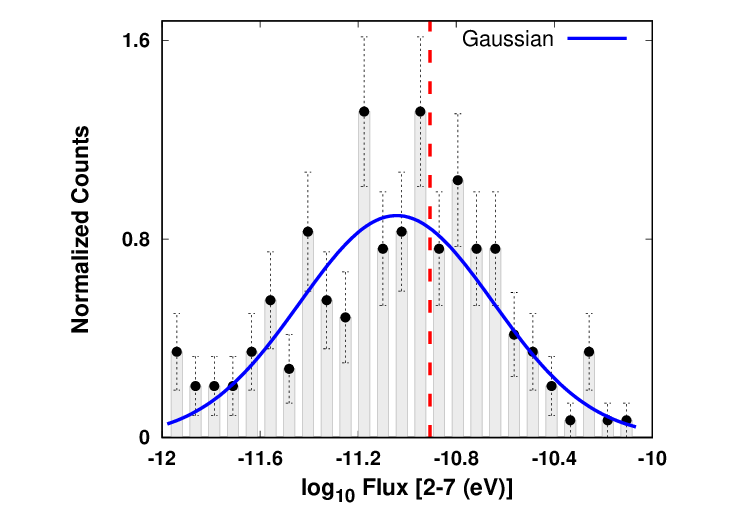}
		\caption{}
		\label{fig:a}
	\end{subfigure}
	\begin{subfigure}{0.49\textwidth}
		\includegraphics[scale=0.74]{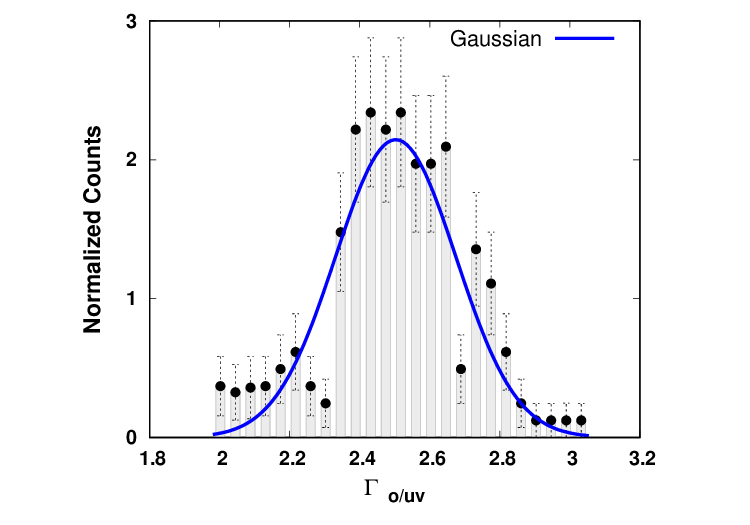}
		\caption{}
		\label{fig:b}
	\end{subfigure}

	\caption{Histograms of logarithmic flux and index distributions of optical/UV spectrum of 3C\,279. The solid curve represents the best fitted single Gaussian function represented by equation \ref{eq:gauss}. The dotted vertical line in the left panel gives the value of mean flux.}

\label{fig:hist_ouv}

\end{figure*}

\begin{figure*}
	\begin{subfigure}{0.49\textwidth}
		\includegraphics[scale=0.74]{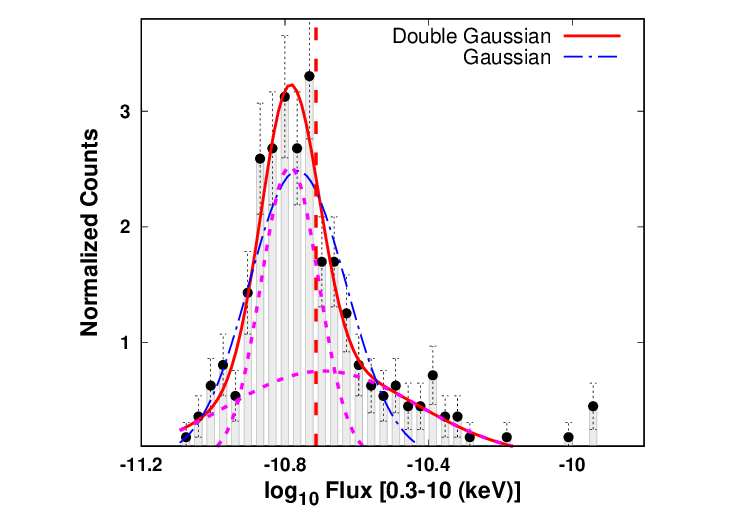}
		\label{fig:a}
	\end{subfigure}
	\hfill
	\begin{subfigure}{0.49\textwidth}
		\includegraphics[scale=0.74]{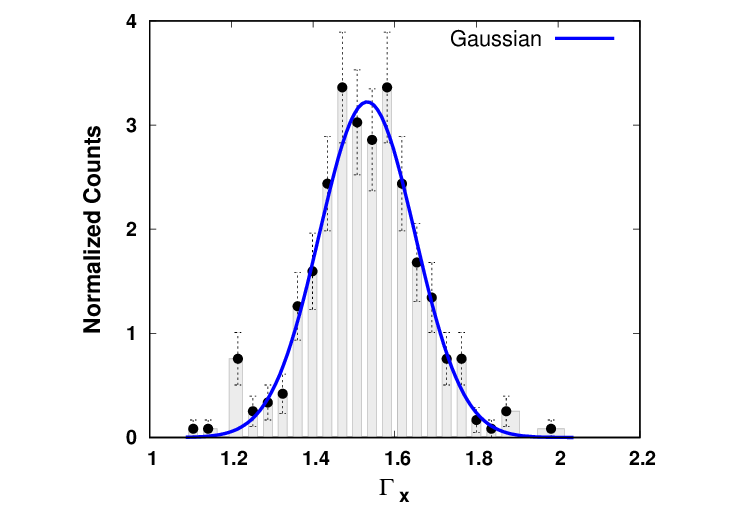}
		\label{fig:b}
	\end{subfigure}

\caption{Histograms of logarithmic flux and index distributions of X-ray emission in 3C\,279. (a): The solid curve represents the best fitted double gaussian function represented by equation \ref{eq:dpdf}, dotted dashed curve gives the fitted single Gaussian (equation \ref{eq:gauss}), dotted curves display the two components of double Gaussian function and dashed vertical line gives the logarithm of average flux value. (b): The solid curve represent the best fitted Gaussian function to the histogram of index.}

\label{hist-x-ray}	
\end{figure*}

We repeated this analysis by studying the histograms of the logarithm of fluxes and the spectral indices. 
The histograms are 
then fitted with Gaussian and double Gaussian probability density functions (PDF). A Gaussian PDF is defined as
\begin{equation}\label{eq:gauss}
	\textbf
	\rm f(x) = \frac{a}{\sqrt{2\pi \sigma^2}} \,\,e^\frac{-(x-\mu)^2}{2\sigma^2}
\end{equation}
where, $\mu$  and $\sigma$ are the mean and standard deviation of the distribution; while a double Gaussian PDF is defined as 
\begin{equation}\label{eq:dpdf}
\rm d(x) = \frac{a}{\sqrt{2\pi \sigma_1^2}}\,\, e^\frac{-(x-\mu_1)^2}{2\sigma_1^2} \
+ \frac{1-a}{\sqrt{2\pi \sigma_2^2}}\,\, e^\frac{-(x-\mu_2)^2}{2\sigma_2^2} 
\end{equation}
where $\mu_1$, $\mu_2$, $\sigma_1$ and $\sigma_2$ are the means and standard deviations of the two Gaussian PDFs.
Consistent with the AD test, histograms of the logarithm of optical/UV fluxes and the corresponding spectral indices can be well
fitted by a Gaussian PDF (with reduced chi-squares $\chi_{\rm red}^2 = 1.08$ and $\chi_{\rm red}^2 = 1.28$ respectively). In Figure \ref{fig:hist_ouv}, 
we show these histograms with the best fit Gaussian PDF. For X-rays, the fit by a double Gaussian PDF 
to the histogram of the logarithm of X-ray fluxes provided better statistics ($\chi_{\rm red}^2 = 0.86$) compared to Gaussian PDF 
($\chi_{\rm red}^2 = 1.99$); whereas, the Gaussian PDF was able to fit the X-ray spectral index histogram successfully. Though the 
logarithm of X-ray fluxes favoured a double Gaussian PDF, the distributions are too close to be differentiated. In
Figure \ref{hist-x-ray}, we show these distributions with the best-fit Gaussian (logarithm of fluxes and indices) and the double Gaussian
(logarithm of fluxes) PDFs.

\begin{figure*}
	\begin{subfigure}{0.49\textwidth}
		\centering
		\includegraphics[scale=0.74]{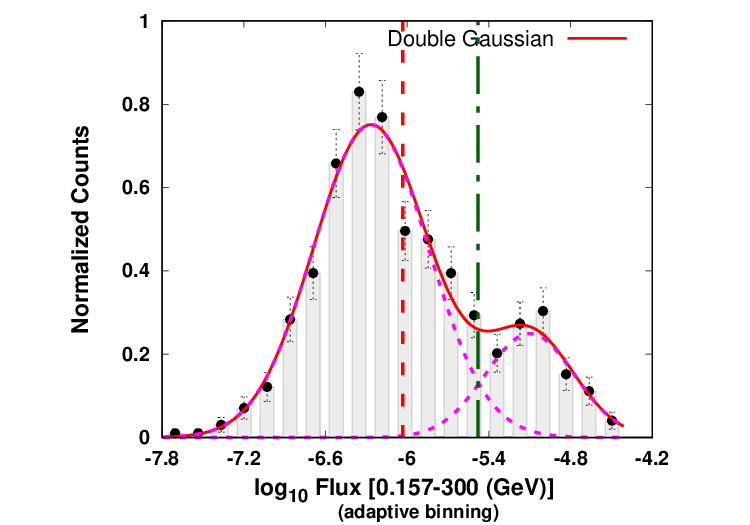}
		\label{fig:a}
	\end{subfigure}
	\hfill
	\begin{subfigure}{0.49\textwidth}
		\includegraphics[scale=0.74]{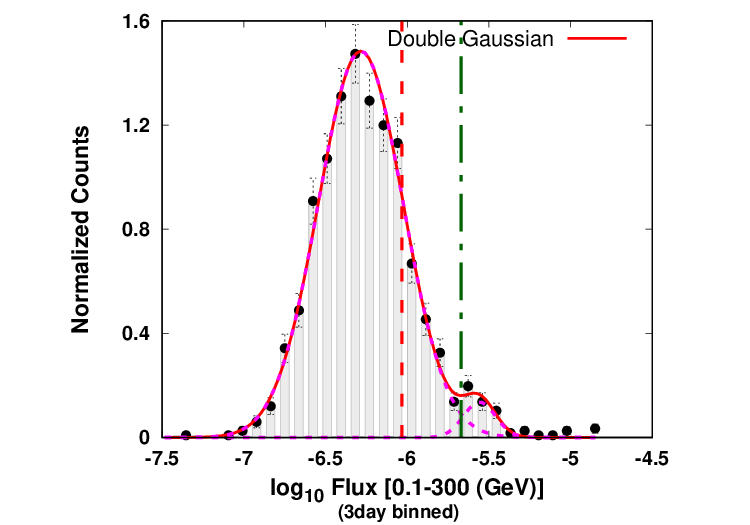}
		\label{fig:a}
	\end{subfigure}
	
	\caption{Plots showing histogram fitting of logarithmic flux distributions of adaptively binned (a)  and 3 day binned (b) Fermi lightcurve. The red solid line represents the best fitted double  Gaussian function represented by equation\ref{eq:dpdf}.The dotted curves represents the two individual components of the double Gaussian. The vertical dashed line represents the logarithm of average flux value and the dotted dashed vertical line mark the point of intersection of two possible flux states for the $\gamma$-ray emission. The adaptively binned fluxes are estimated in 0.157-300 GeV energy range.} 

\label{hist-gamma-ray1}	
\end{figure*}

\begin{figure*}
	\begin{subfigure}{0.49\textwidth}
		\centering
		\includegraphics[scale=0.74]{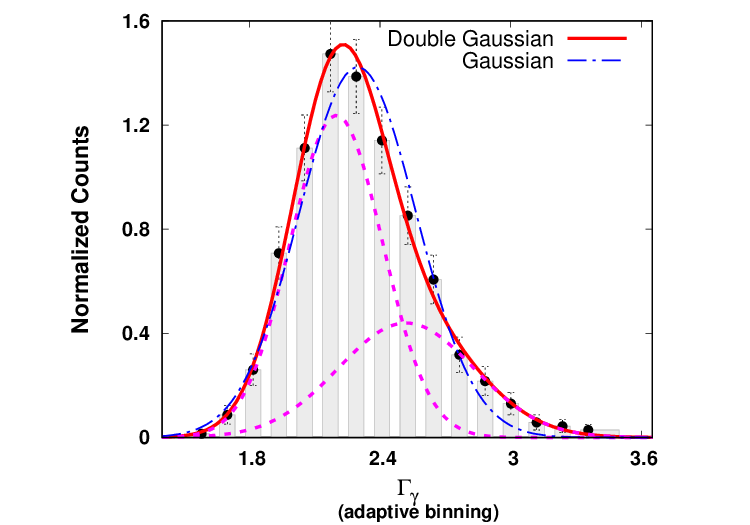}
		\label{fig:b}
	\end{subfigure}
	\hfill
	\begin{subfigure}{0.49\textwidth}
		\includegraphics[scale=0.74]{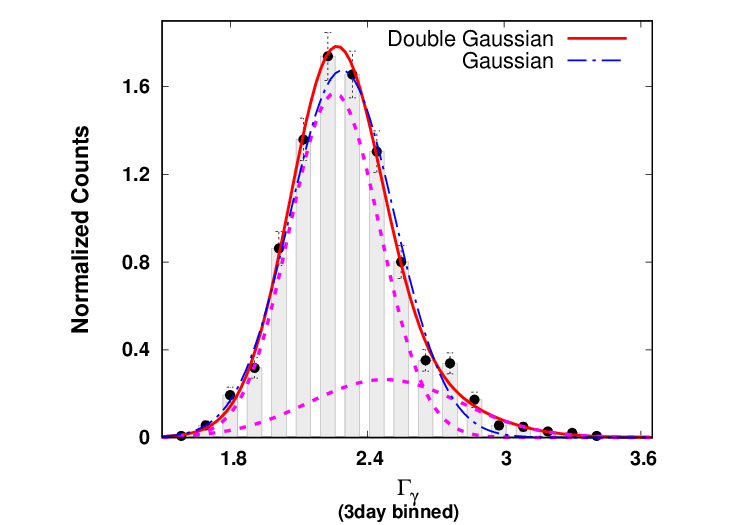}
		\label{fig:b}
	\end{subfigure}
	
\caption{Histogram fitting of Index distributions of adaptively binned (a)  and 3 day binned (b) Fermi lightcurves. The solid curve represents the best fitted double  Gaussian function and dotted dashed curve gives the best fitted single Gaussian. The dotted curves display the two components of the double Gaussian.} 

\label{hist-gamma-ray2}	
\end{figure*}

The histogram of the logarithm of $\gamma$-ray fluxes constructed from the adaptively binned light curve clearly 
showed two distinct peaks and is well fitted by a double Gaussian PDF ($\chi_{\rm red}^2=0.824$). A similar feature is 
also observed in 3-day binned fluxes and the best fit double Gaussian PDF resulted in similar statistical moments 
(means and the standard deviations). The results are given in Table \ref{tab:dist} and the histograms with best-fit double Gaussian PDFs are shown in Figure \ref{hist-gamma-ray1}. 
In both cases, the index distribution also favoured double Gaussian
PDFs and the fit statistics are given in Table \ref{tab:dist}. In Figure \ref{hist-gamma-ray2}, we show the corresponding histograms and the best fit
Gaussian and double Gaussian PDFs. Though the distribution of two Gaussians of the double Gaussian PDF is not 
very distinct in the histogram of indices, they supplement the double Gaussian flux distribution with the fact that 
the plausible log-normal $\gamma$-ray flux variability can be associated with the Gaussian variability in indices.
However, to affirm this claim more rigorous statistical treatment will be required. In the histogram of logarithm of 
fluxes, we demarcate the mean $\gamma$-ray flux by vertical dashed lines. Based on these histograms, the average flux
may not actually demarcate the high and low states instead one can use the flux value at which the two Gaussian functions
intersect (dotted dashed vertical line in Figures \ref{hist-gamma-ray1} (a) and (b)). However, we encountered very few observations during which the 
fluxes are higher than this demarcated value ($3.3\times 10^{-6}$ phs cm$^{-2}$ s$^{-1}$) and hence did not use in this work.

\begin{table*}
\setlength\extrarowheight{3pt}
\setlength\tabcolsep{2pt}
\scriptsize
\begin{tabular}{c c c c c c c c c c }
	
	\hline
	\hline
	& Histogram  & PDF & $\mu_1$ & $\sigma_1$  & $\mu_2$ & $\sigma_2$  &     \multicolumn{1}{c}{$a_1$}  &   dof & $\rm \chi^2/dof$ \\
	\hline

	optical/UV  & log10(Flux) & Gaussian &-11.04 $\pm$   0.03  & 0.39$\pm$0.03  &  &   & 0.89 $\pm$ 0.07  & 22 & 1.08\\
	& Index  & Gaussian & 2.50$\pm$0.01 & 0.17$\pm$0.01 & &  & 0.92$\pm$ 0.08 &   18 & 1.28\\
	
	X-ray & log10(Flux) & double Gaussian& -10.786 $\pm$   0.01  & 0.08$\pm$0.01  & -10.69$\pm$0.03 & 0.26$\pm$0.03 & 0.50$\pm$0.07 & 25 & 0.86 \\
	& & Gaussian &-10.76 $\pm$ 0.01 & 0.13  $\pm$ 0.01 & & & 0.84 $\pm$ 0.07 &  27 & 1.99\\ 
	& Index  & Gaussian &1.53$\pm$0.006 & 0.12$\pm$0.005 & &  & 0.94$\pm$ 0.05 &   19 & 0.975\\
	
	3 day binned  & log10(Flux) & double Gaussian & -6.284$\pm$0.01 & 0.259$\pm$0.01  & -5.557$\pm$0.02 & 0.097$\pm$0.02 & 0.967$\pm$0.01& 22 & 1.14\\
	& Index  & double Gaussian &2.256$\pm$0.01 & 0.198$\pm$0.01 & 2.476$\pm$0.09 & 0.331$\pm$0.03 & 0.781$\pm$0.11 & 14 & 1.31\\
	
	& & Gaussian & 2.289 $\pm$ 0.01 & 0.229$\pm$0.01 & & & 0.962$\pm$0.05 & 16& 3.25\\
	
	Adaptive binning  & log10(Flux) & double Gaussian & -6.266$\pm$0.03 & 0.423$\pm$0.02  & -5.098$\pm$0.06 & 0.324$\pm$0.05 & 0.783$\pm$0.04 & 14 & 0.824 \\
	
	

	& Index  & double Gaussian &2.199$\pm$0.02 & 0.211$\pm$0.03 & 2.519$\pm$0.28 & 0.310$\pm$0.08 & 0.657$\pm$0.35 & 48 & 1.42\\
	&   & Gaussian & 2.297$\pm$0.01 & 0.268$\pm$0.01 & && 0.956$\pm$0.05 & 14&1.90\\
	
	\hline
	\hline
	
\end{tabular}
\caption{Best fit parameter values of the PDF functions  fitted to the logarithm of flux and index histograms for X-ray, optical/U and $\gamma$-ray light curves.}
\label{tab:dist}

\end{table*}



\section{Broadband SED Analysis}
\label{broadband}
\begin{figure*}
	\begin{subfigure}{0.45\textwidth}
		\hspace{-5mm}
		\includegraphics[width=0.75\textwidth,angle=-90]{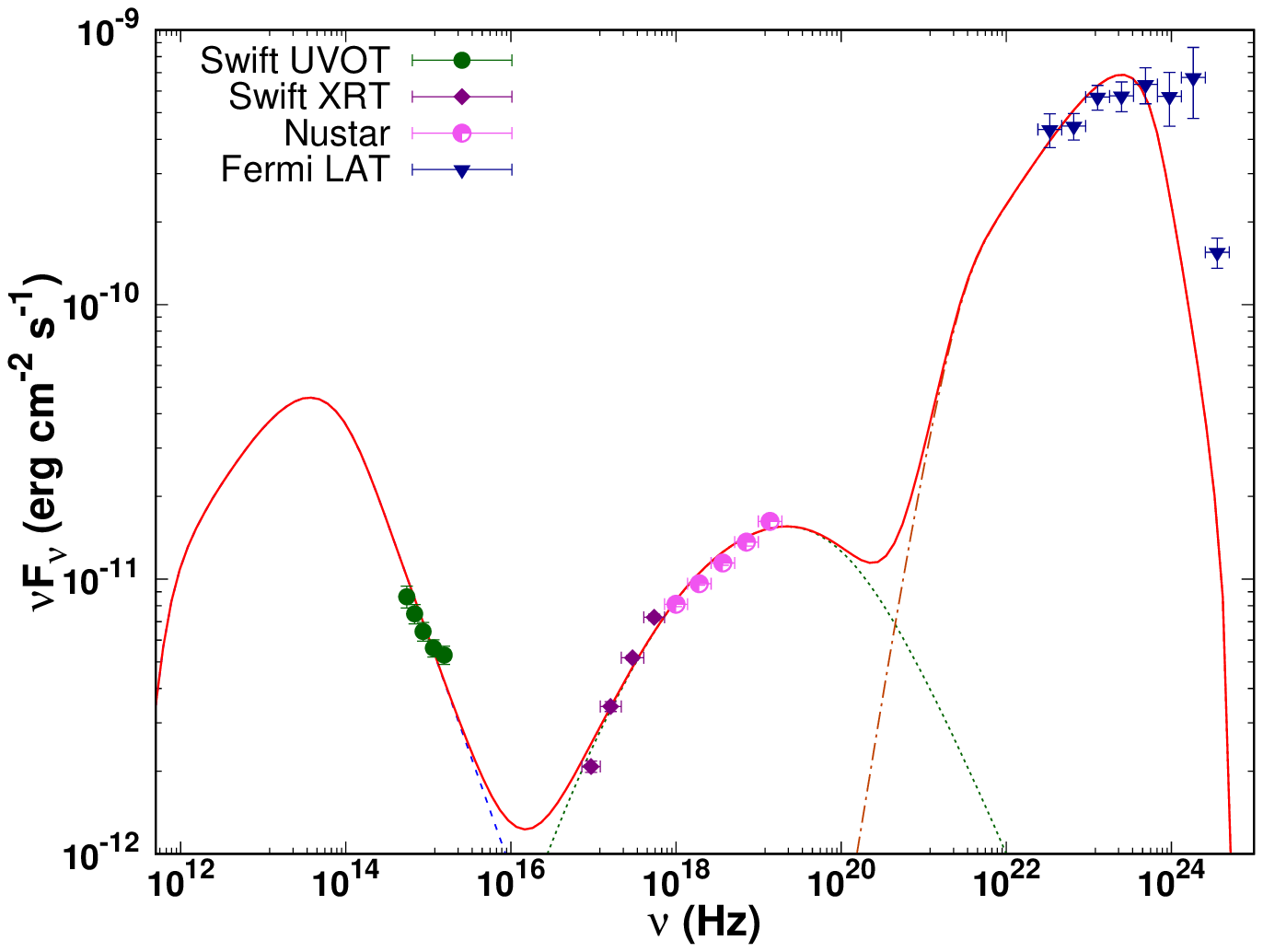}
		\label{fig:a}
	\end{subfigure}
	\hspace{-4mm}
	\begin{subfigure}{0.45\textwidth}
		\includegraphics[width=0.7\textwidth,angle=-90]{sed1.eps}
		\label{fig:b}
	\end{subfigure}
	\caption{Figure showing the broadband SED fitting of 3C\,279 during the time interval 56642-56649 considering Synchrotron, SSC and EC emissions. \textbf{Left}: Dashed curve represents the synchrotron, dotted curve shows SSC, and dashed dotted curve represents the EC-BLR components respectively. The solid curve shows the combined model. The value of $\chi^{2}$ for this fit is 27 out of 17 degrees of freedom. \textbf{Right}: The plot of unfolded spectrum with residuals generated form XSPEC.}
	\label{sed1}
	
\end{figure*}

\begin{figure*}
	
	\begin{subfigure}{0.48\textwidth}
		\hspace{-5mm}
		\includegraphics[width=0.75\textwidth,angle=-90]{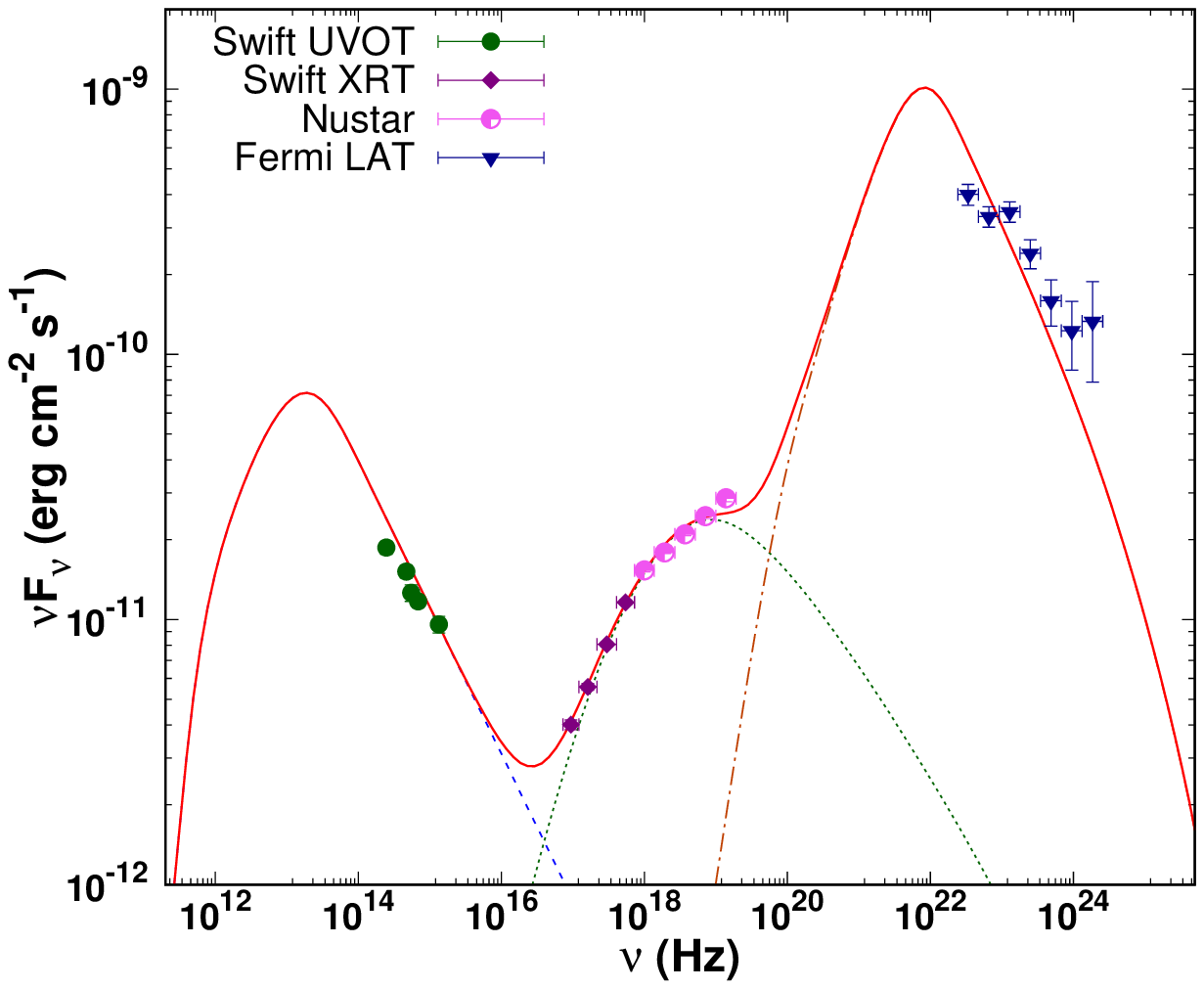}
		\label{fig:a}
	\end{subfigure}
	\hspace{-4mm}
	\begin{subfigure}{0.45\textwidth}
		\includegraphics[width=0.7\textwidth,angle=-90]{sed2.eps}
		\label{fig:b}
	\end{subfigure}
	\caption{Figure showing the broadband SED fitting of 3C\,279 during the time interval 56649-56660 considering Synchrotron, SSC and EC emissions. \textbf{Left}: Dashed curve represents the synchrotron, dotted curve shows SSC, and dashed dotted curve represents the EC-IR components respectively. The solid curve represents the sum of all components. The value of $\chi^{2}$ for this fit is 23.97 out of 16 degrees of freedom. \textbf{Right}: The plot of unfolded spectrum with residuals generated form XSPEC.}
	\label{sed2}
	
\end{figure*}

\paragraph*{}
The multi-wavelength correlation analysis between different spectral quantities (\S\ref{correlations}) suggests the optical/UV emission
fall on the high energy end of the synchrotron SED. The X-ray and $\gamma$-ray energy bands fall on the Compton SED
and the linear regression analysis suggests different emission processes to be active at these energies. Further, the 
$\gamma$-ray energy band falls around the peak of the Compton spectral component. To validate these findings, we performed
a detailed spectral modelling of the simultaneous observation of the source using synchrotron, SSC and EC processes. For this,
we selected two epochs with simultaneous observation by \emph{Swift}, \emph{Nu}STAR, and \emph{Fermi} telescopes. This provides
a wealth of information for the broadband spectral fitting and the resultant observed SED is shown in Figures \ref{sed1} and \ref{sed2}.

To model the SED, we consider a spherical emission region of radius $R$ moving down the blazar jet relativistically with 
Lorentz factor $\Gamma$ at an angle $\theta$ with respect to the line of sight. The emission region is populated with a
relativistic broken power-law electron distribution described by
\begin{align} \label{eq:broken}
	N(\gamma)\,d\gamma = \left\{
	\begin{array}{ll}
		K\,\gamma^{-p}\,d\gamma&\textrm{for}\quad \mbox {~$\gamma_{\rm min}<\gamma<\gamma_b$~} \\
		K\,\gamma_b^{q-p}\gamma^{-q}\,d\gamma&\textrm{for}\quad \mbox {~$\gamma_b<\gamma<\gamma_{\rm max}$~}
	\end{array}\quad {\rm cm}^{-3}
	\right.
\end{align}
Here, $\gamma$ is the electron Lorentz factor with $\gamma_{\rm min}$ and $\gamma_{\rm max}$ are the minimum and the maximum
Lorentz factor of the electron distribution, $p$ and $q$ are the low and high energy indices of the distribution with 
$\gamma_b$ is the Lorentz factor corresponding to the break, and $K$ is the normalization. The emission
region is permeated with a tangled magnetic field, $B$ and the electron distribution loses its energy through synchrotron,
SSC, and EC processes. Depending upon the location of the emission region from the central black hole, the dominant external
photon field can be either Ly-$\alpha$ line emission from the BLR (EC/BLR) or the thermal IR photons (EC/IR) from the dusty 
torus \citep{2009MNRAS.397..985G}. The emissivities due to these radiative processes are estimated numerically and the observed 
flux at earth is obtained after considering the relativistic and 
cosmological effects \citep{1995ApJ...446L..63D, 1980Natur.287..307B}. For numerical simplicity, we treated 
the BLR photon field as thermal distribution corresponding to temperature 42000 K (equivalent to Ly-$\alpha$ line frequency)
and the temperature of the IR photon field was chosen as 1000 K \citep{2018RAA....18...35S,2000ApJ...545..107B}.

The numerical model which is capable of producing the synchrotron, SSC, and EC spectrum from the source parameters was 
added as a local model in XSPEC and used to
fit the broadband SED corresponding to the selected two epochs (56642--56649 MJD and 56649--56660 MJD). 
To reduce the number of free parameters, we assumed
equipartition between the magnetic field and the electron energy densities \citep{1959ApJ...129..849B,1999qagn.book.....K}. 
The proton population, responsible for providing the bulk jet power, are assumed to be cold in the reference frame of the emission 
region and hence was not included in deriving the equipartition condition. 
The limited information available at optical/UV, X-ray and $\gamma$-ray energies do not let us to constrtain all 
the parameters and hence,\,the fitting was performed 
only on four parameters namely, $p$, $q$, $\Gamma$ and $B$. The rest of the parameters are frozen at their typical 
values or at a value that provides better fit statistics. The values chosen for the frozen parameters are $\theta=2\deg$.
$\gamma_{\rm min}=40$, $\gamma_{\rm max}=10^{7}$, and $R\sim 10^{16}$. The temperature of the external photon field is 
frozen either at 1000 K (EC/IR) or 42000 K (EC/BLR) depending upon the fit statistics. The parameter $\gamma_b$ can be constrained with the 
knowledge of the  
photon frequencies at which synchrotron and SSC spectral component peaks \citep{2018RAA....18...35S}.
\begin{align}
	\gamma_b = \sqrt{\frac{\nu_{\rm{ssc,peak}}}{\nu_{\rm{syn,peak}}}}
\end{align}
 However, the available information at optical/UV and X-ray energies do not let us to obtain these peak 
frequencies. Nevertheless, since $\Gamma_{\rm o/uv} > 2$ and $\Gamma_x < 2$, this suggests that $\nu_{\rm{syn,peak}}$ < 2 eV 
and $\nu_{\rm{ssc,peak}}$ > 79 keV corresponding to minimum photon energy of optical observation 
and maximum photon energy of X-ray observation.
A constraint on $\gamma_b$ can then be obtained as $\gamma_b > 200$. 
The initial spectral fit was performed with $\gamma_b$ setting as a free parameter and satisfying the above constraint.
We find that the choice of $\gamma_b$ as 750 for the epoch 56642--56649 MJD and 400
for the epoch 56649--56660 MJD provide better fit statistics. The fit is repeated with $\gamma_b$ frozen at 
these values and the results are 
given in Table \ref{tab:sed}. 
In order to obtain the fit statistic $\chi_{\rm red}^2 <2$, we included additional 12\% of systematic error evenly to all the 
data. This criterion was necessary for the XSPEC to provide the confidence intervals on the best fit parameters.

The $\gamma$-ray spectrum corresponding to the epoch 56642--56649 MJD is hard and this demanded the EC peak to fall at 
high energies to obtain a better fit. Consistently, we found EC/BLR provides a better fit to the SED with 
$\chi_{\rm red}^2 =1.58$ as compared to EC/IR ($\chi_{\rm red}^2 > 10$). On the contrary, the $\gamma$-ray spectrum
corresponding to the epoch 56649--56660 MJD is relatively soft and indicate the EC peak frequency falls at lower 
energies. Through the SED fitting, we also noted that EC/IR interpretation of the $\gamma$-ray spectrum 
provides a better fit statistic ($\chi_{\rm red}^2 =1.5$) than EC/BLR ($\chi_{\rm red}^2 > 10$) for this epoch. 
It can be argued
that with a proper choice of $\gamma_b$, the $\gamma$-ray spectrum during these epochs can be explained 
under single emission process (EC/IR or EC/BLR). However, this was not favoured 
by our initial fit to the SED with $\gamma_b$ set as a free 
parameter. We found better fit statistics when $\gamma_b$ is approximately equal to the values 
quoted in Table \ref{tab:sed} along with mentioned target photon fields.  
Hence, the SED fitting
for these two epochs indicates a significant variation in the target photon field (IR from the dusty environment
and the Ly-$\alpha$ emission from BLR) and this suggests 
during different flaring epochs, the emission region may be located at different distances from the central
black hole \citep{2009MNRAS.397..985G}. 
Interestingly, this analysis implies that the variation in the EC spectral peak of 
3C\,279 is associated with the change in the frequency of external photon field.
To be consistent with 
our inference obtained from the correlation study (\S\ref{trans-corre}), where 
the variation in Compton SED peak is attributed to the changes in $\gamma_b$,
we interpret that the variations in Compton peak 
may be associated with the change in both $\gamma_b$ and the external photon field.
The spectral fit results presented here may vary moderately, if we relax the assumption of equipartition condition 
which banks upon the minimum total energy criterion of the source \citep{1959ApJ...129..849B,1999qagn.book.....K}. 
However, this do not affect our conclusion on the variation in the target photon field for the EC process. The reason for this being, 
the equipartition condition constrains the particle and magnetic field energy densities; whereas, the EC spectral
component depend upon the particle distribution and the external photon field with no direct dependence on the 
jet magnetic field \citep{2017MNRAS.470.3283S}.
\begin{table*}
	\caption{Best fit values of  the source parameters from broadband SED fitting.}
 	\label{tab:sed}
 	\setlength{\tabcolsep}{15pt}
 	\begin{tabular}{llll}
 		\hline\hline
 		Name of parameter& Symbol & 56642-56649 & 56649-56660  \\
 		\hline\hline
 	Low energy Particle index    &  p      &    1.961$\pm$0.13 &    1.07$\pm$0.27    \\
 	High energy Particle index    &     q    &    4.687$\pm$0.11  &    4.103$\pm$0.10   \\
 	Bulk Lorentz factor& $\Gamma$ & 14.05$\pm$0.79 & 17.62$\pm$1.5 \\
 	Magnetic Field (G)  & B & 1.598$\pm$0.04 & 1.52$\pm$0.03 \\
 	External Compton Process & &EC/BLR & EC/IR(1000K) \\
 	External photon energy density (erg/cm$^3$)&U$_*$ & $7.06\times10^{-3}$ & $2.269\times10^{-3}$ \\
 	Total jet power & P$_{\rm jet}$ & $4.47\times10^{45}$ & $4.72\times10^{45}$ \\
 	Total radiated power & P$_{\rm rad}$ & $1.403\times10^{43}$ & $1.303\times10^{43}$  \\
 	\hline\hline
 \end{tabular}
 \end{table*}

\section{Summary}
\label{sec:summary}
\paragraph*{}
We performed a detailed long term spectral analysis of the FSRQ 3C\,279 using simultaneous broadband observations 
of the source by \emph{Swift} XRT, UVOT, and \emph{Fermi}-LAT observations.
The spectra at these energy bands can be individually fitted by a power-law and we found a clear ``harder when brighter''
trend at X-ray energy; however, no such behaviour was witnessed in optical/UV and $\gamma$-ray energies. We also 
estimated the transition photon energy at which the dominance of the synchrotron and the inverse Compton spectral
component switches. The transition energy was well correlated with the optical/UV flux and the X-ray spectral index
but not with the other quantities.
This correlation study suggests, at X-ray energies the flux enhancement is mainly dominated by the variations in the
spectral index while the change in normalization is associated with the flux variations at optical/UV energies.
We also find a moderate correlation between the optical/UV and X-ray spectral indices; however, the linear regression
analysis disfavoured the radiative loss origin of the broken power-law electron distribution responsible for the
broadband emission. These study results let us conclude that multiple acceleration process may be responsible
for the broken power-law electron distribution and the long term spectral variations are predominantly associated with the 
power-law index changes at lower energy ($E<E_b$) and the variation in break energy at higher energy ($E>E_b$). This is also consistent with the range of $\gamma$-ray spectral indices.

The long term flux and index distribution of the source were also studied to identify the nature of variability.
The Anderson-Darling test suggested the optical/UV flux is exhibiting a lognormal distribution and the corresponding index distribution showed
normal behaviour. Further, the $\gamma$-ray flux distribution clearly showed a double log-normal feature 
consistent with double Gaussian behaviour of the spectral index distribution. These results are consistent with log-normal 
variability of the source suggested by the earlier studies. However, the flux variations may be associated with the 
changes in the index rather than highlighting the link between the blazar jet and the accretion disk.

The broadband SED of the source was performed considering synchrotron, SSC, and EC processes for two epochs with 
simultaneous observations by \emph{Swift}, \emph{Nu}STAR, and \emph{Fermi} telescopes. Our fitting result suggests,
optical/UV emission is associated with the synchrotron emission process while the X-ray and $\gamma$-ray emissions
are due to SSC and EC processes. Among the two selected epochs, the one with hard $\gamma$-ray spectrum (56642-56649 MJD) indicates
EC/BLR origin for the $\gamma$-ray emission and the epoch with soft $\gamma$-ray spectrum (56649-56660 MJD) supports EC/IR
emission process. Interestingly, these SED fit suggest the variations in $\gamma$-ray peak may be associated with 
the change in target photon frequency. Combining this result with the conclusions drawn from the correlation
study, we find the variation in the peak of the $\gamma$-ray spectral component can be associated with the changes
in the break energy of the electron distribution and the temperature of the target photon field involved in the 
EC process.
\section*{Acknowledgements}
\paragraph*{}
We thank the referee Narek Sahakyan for the valuable suggestions which helped to improve the work
significantly.
This research work has made use of data obtained from NASA's High Energy Astrophysics 
Science Archive Research Center (HEASARC) and Fermi gamma-ray telescope Support centre, a service of the Goddard Space Flight Center and the Smithsonian Astrophysical Observatory.
AT is thankful to UGC-SAP and FIST 2 (SR/FIST/PS1-159/2010) (DST, Government of India) 
for the research facilities provided in the Department of Physics, University of Calicut.
SZ is supported by the Department of Science and Technology, Government of India, under the INSPIRE Faculty grant, DST/INSPIRE/04/2020/002319.
\section*{Data Availability}
The data used in this work are publicly available and downloaded from the
archives at \url{https://heasarc.gsfc.nasa.gov/} and \url{https://Fermi.gsfc.nasa.gov/}.
\bibliographystyle{mnras}
\bibliography{paper}
\bsp	
\label{lastpage}


\end{document}